# Low temperature equilibrium correlation functions in dissipative quantum systems

Stefan K. Kehrein[1] and Andreas Mielke[2]

Institut für Theoretische Physik,
Ruprecht–Karls–Universität,
Philosophenweg 19,
D-69120 Heidelberg, F.R. Germany

November 18, 2017

## Abstract

We introduce a new theoretical approach to dissipative quantum systems. By means of a continuous sequence of infinitesimal unitary transformations, we decouple the small quantum system that one is interested in from its thermodynamically large environment. This yields a trivial final transformed Hamiltonian. Dissipation enters through the observation that generically observables "decay" completely under these unitary transformations, i.e. are completely transformed into other terms. As a nontrivial example the spin–boson model is discussed in some detail. For the super–Ohmic bath we obtain a very satisfactory description of short, intermediate and long time scales at small temperatures. This can be tested from the generalized Shiba–relation that is fulfilled within numerical errors.

---

[1]E–mail: kehrein@marvin.tphys.uni-heidelberg.de
New permanent address: Theoretische Physik III, Institut für Physik, Universität Augsburg, D-86135 Augsburg, F.R. Germany

[2]E–mail: mielke@hybrid.tphys.uni-heidelberg.de

# Contents

# 1 Introduction

## 1.1 Methods for dissipative quantum systems

Dissipative quantum systems, i.e. quantum systems coupled to an environment, occur in many areas of physics and chemistry and have been studied extensively in the past using various methods. In their pioneering work Caldeira and Leggett [1, 2] proposed a standard model of dissipative quantum systems describing a particle coupled to a thermodynamically large environment modelled by harmonic oscillators. The typical Hamiltonian for such a problem is of the form

$$H = H_S + H_B + H_{SB}. \qquad (1.1)$$

$H_S$ is the Hamiltonian of the system, $H_B$ describes the environment and $H_{SB}$ is the coupling between system and environment. A recent review of methods in the field of dissipative quantum systems can be found in Ref. [3].

Most approaches start off by discussing the time evolution of the reduced density matrix of the small quantum system. This is achieved by eliminating the bath degrees of freedom, for example by taking a partial trace or by integrating them out in a path integral framework using the Feynman–Vernon influence functional method [4]. This yields some effective action with interactions that are non–local in imaginary time [5]. For a general dissipative quantum system some approximations become necessary in this effective action, for example the NIBA (Non–Interacting Blip Approximation) in the spin–boson model [5].

Another important approach is the description as a quantum Markov process. However, from a fundamental point of view this description should be derived from an underlying system plus bath Hamiltonian. It is well–known that this is only possible in certain limiting cases, for a careful discussion see e.g. [6].

Our approach introduced in this work avoids this problem altogether by staying in a Hamiltonian framework. A well–developed method for Hamiltonians is provided by renormalization theory. Three decades ago Wilson [7] studied the problem of a fixed source Hamiltonian using renormalization. This model is a simple dissipative quantum system. A nucleon with infinite mass, the fixed source, is coupled to $\pi$ mesons. The nucleon contains only an internal degree of freedom, it can be a proton or a neutron. It is described by a two–level system. The two–level system is coupled linearly to two bosonic baths describing the two $\pi$ mesons. Wilson analysed this model using non–perturbative renormalization. A similar renormalization scheme has been applied to the problem of dissipative tunneling by Bray and Moore [8] and by Chakravarty [9]. In the simplest case of dissipative tunneling one considers a particle in a double–well potential coupled to the bosonic excitations of the environment. The particle in the double–well potential can be approximated by a two–level system. In the review of Leggett et al. [5] the adiabatic renormalization scheme applied to the dissipative two–level system is also discussed.

In a renormalization scheme the Hamiltonian of system plus bath is mapped to an effective Hamiltonian that contains no high energy modes but has the same low energy spectrum as the original Hamiltonian. The main problem is that the renormalized Hamiltonian still contains a large part of the bosonic modes of the environment, namely the slow and the resonant modes. Therefore it is rather difficult to calculate time–dependent correlation functions using this renormalized Hamiltonian. In fact no example is known to the authors where such a renormalized Hamiltonian has been explicitly useful beyond the identification of the low–energy scale. Alternatively one can use numerical renormalization [10]. But for dissipative quantum systems this method has (at least until today) a restricted range of applicability: One can only treat fermionic baths with this method. Therefore one has to find a mapping from the problem of interest with a bosonic bath to a model with a fermionic bath. This is only possible if the bosonic bath has special spectral properties. Only recently this approach was used to calculate the dynamical properties of the dissipative two–level system [11], which can be mapped to the Kondo problem if one starts off with an Ohmic bath.

Our approach used in this paper can be seen as an extension of traditional renormalization methods. By decoupling states with respect to their specific energy differences (and not treating high–energy modes only starting from the UV–cutoff), we are finally left with a simple effective Hamiltonian. This Hamiltonian contains only nearly resonant modes and can at zero temperature be solved without problems. Let us mention that a similar renormalization scheme has recently also been introduced by Glazek and Wilson in high–energy physics [12]. Obviously our approach is fundamentally different from the usual methodology as we do not employ a reduced density matrix formalism. We rather obtain a final Hamiltonian where the small quantum system and the environment are decoupled.

## 1.2 Flow equations for Hamiltonians

Let us explain the approach used in this paper in more detail. Technically it is based on a continuous sequence of infinitesimal unitary transformations that is applied to the Hamiltonian. This technique has been proposed by Wegner to bring a given many–particle Hamiltonian in diagonal or block–diagonal form [13]. The continuous sequence of unitary transformations $U(\ell)$ is labelled by a flow parameter $\ell$ with dimension (Energy)$^{-2}$. It corresponds to the energy difference that is just being decoupled, that is for small $\ell$ large energy differences are decoupled first before smaller energy differences are dealt with later for larger values of $\ell$.

Applying such a transformation to a given Hamiltonian, this Hamiltonian (i.e. its parameters) become functions of $\ell$. Usually it is more convenient to work with a differential formulation

$$\frac{dH}{d\ell} = [\eta(\ell), H(\ell)], \quad H(\ell = 0) = H \tag{1.2}$$

with an anti–hermitian generator $\eta(\ell)$ related to the unitary transformation by

$$\eta(\ell) = \frac{dU^\dagger(\ell)}{d\ell} U(\ell). \tag{1.3}$$

With a suitable choice of $\eta(\ell)$ and additional approximations which will be described later, one is able to solve the differential equation (1.2). Thereby one obtains a final Hamiltonian $H(\ell = \infty)$. The necessary approximations are chosen so that the low energy spectrum of the initial Hamiltonian is well–reproduced by the final Hamiltonian $H(\ell = \infty)$. In previous work [14, 15] we have already used this method to obtain the long–time behaviour of equilibrium correlation functions in the spin–boson model. In the present paper we show how the method can be improved to obtain information about equilibrium correlation functions also for intermediate ($t \Delta_r \approx 1$) and short times ($t \Delta_r \ll 1$).

The generator $\eta(\ell)$ is chosen so that the initial Hamiltonian (1.1) is mapped onto a final Hamiltonian which has the simple structure

$$H(\ell = \infty) = H_{S\infty} + H_B. \tag{1.4}$$

The coupling $H_{SB}$ is eliminated. $H_{S\infty}$ is a renormalized or effective Hamiltonian of the quantum system and $H_B$ is the Hamiltonian of the environment. The latter remains unchanged since we assume that the bath is large compared to the system and therefore does not change its properties when it is coupled to the small system. Having such a procedure in mind two questions are immediate:

- At first sight it is quite unclear how a Hamiltonian like (1.4) can describe dissipation. Since the small system is decoupled from the environment, exchange of energy between system and environment is no longer possible. To resolve this question, one has to understand how time–dependent correlation functions are calculated within our formalism. Usually dissipation manifests itself in a decay of correlation functions like $\langle q(t) q(0) \rangle$, where $q$ is the coordinate of the particle coupled to the bath. We will later see that in our approach dissipation enters through the following simple fact: If one performs a unitary transformation, all the observables have to be transformed as well. In the case of a single particle, one has for example to transform the coordinate $q$. This means that one has to solve the flow equation

$$\frac{dq}{d\ell} = [\eta, q] \qquad (1.5)$$

Transforming an observable of the system, it will finally contain many environment operators. Typically the transformed coordinate $q(\ell = \infty)$ will be of the form $q(\ell = \infty) = \sum_k c_k q_k$, where $q_k$ are coordinates of the bath oscillators and $c_k$ are some numbers. In order to calculate correlation functions we have to use the transformed observables and the transformed Hamiltonian. In this way it will be shown that the information on dissipation is contained in the unitary transformation that has been used to decouple the system from the environment. This will become clear in the special examples we discuss in Sects. 3 and 4.

- Which choice of $\eta$ is appropriate for the problem we have in mind? The main idea is that $\eta$ is chosen in such a way that couplings between system and bath which correspond to large energy differences are eliminated first, couplings which correspond to small energy differences are eliminated later. This yields the fundamental separation of energy scales underlying renormalization theory. A suitable choice of $\eta$ was suggested by Wegner [13] $\eta(\ell) = [H_S(\ell) + H_B(\ell), H_{SB}(\ell)]$. In the present paper we will use a different ansatz, which has some advantages to be discussed later. Nevertheless the couplings between system and bath corresponding to large energy differences are eliminated first.

Another question that arises is whether one really needs infinitesimal unitary transformations. Now in principle it is possible to solve (1.3) and to obtain $U(\ell = \infty)$ as an $\ell$–ordered product of factors $\exp(\eta(\ell) d\ell)$. Since the generator $\eta(\ell)$ does not commute with itself for different values of $\ell$, it is in general not possible to evaluate this infinite product. Of course one could guess some single unitary transformation $U$ right away. But $U(\ell = \infty)$ constructed via the infinitesimal transformations $\eta(\ell)$ has the important energy scale separation automatically built in and this is difficult to ensure if one starts off with one single unitary transformation instead. In fact the unitary transformation constructed in the flow equation approach naturally contains "counter–terms" that eliminate singular induced interaction due to vanishing energy denominators. For the Anderson impurity model this has been discussed in Ref. [16].

Some advantages of this approach are:

+ We work in a Hamiltonian framework.

+ The procedure is non–perturbative as it relies on unitary transformations and has energy scale separation built in. Therefore one finds the correct low–energy scale in the effective Hamiltonian for $\ell \to \infty$.

+ The effective Hamiltonian for $\ell \to \infty$ is simple in the sense that only resonant couplings remain. This is the major advantage as compared to adiabatic renormalization or "poor man's scaling". At low temperature our effective Hamiltonian becomes equivalent to a dissipative harmonic oscillator and equilibrium correlation functions are obtained easily. *The main difference as compared to adiabatic renormalization is that the flow of the observables is not negligible and even very important for $\ell$ corresponding to the low–energy scale.*

Now since one is usually interested in correlation functions of the system, it is sufficient to transform the observables of the system like for example the coordinate and the momentum of a particle. One problem is that apart from special cases like the dissipative harmonic oscillator, the flow equations for the observables are not closed and additional approximations become necessary. In our previous work [14, 15] we have only discussed the long–time behaviour of equilibrium correlation functions of the dissipative two–level system. One goal of the present paper is to show how our method can be improved so that the whole frequency range can be treated. But the transformation of the observables causes also extra problems:

- Since the flow equations for the observables are not closed, a special ansatz for the transformed observables is necessary. In this paper we use an ansatz that is linear in the bosonic modes. Such an ansatz works only for low temperatures, that means $T$ has to be small compared to the typical low–energy scale of $H_S$. At the moment we are not able to obtain reliable results for larger values of $T$ without using traditional approximations of the kind discussed in Sect. 1.1

- Sometimes one is interested in the dynamical behaviour of an observable in a situation, where the system has been prepared in a given initial state. Caldeira and Leggett [2] called this situation *quantum tunnelling* and distinguished it from *quantum coherence*, which is studied in this paper. In our approach we would have to transform the initial state as well. We leave this topic open for future work.

## 1.3 Outline of the paper

The plan of our paper is as follows. In Sect. 2 we develop the general method. A continuous unitary transformation is constructed together with a suitable approximation that decouples the system from the environment. We derive the flow equation for the Hamiltonian describing the quantum system and the flow equation for the spectral function of the bath. We compare our approach with adiabatic renormalization. As a main result we obtain a general expression for the behaviour of dynamical correlation functions at low frequencies. This result is a generalization of our former result for the dissipative two–level system [14]. We show that a dissipative behaviour of correlation functions is obtained only if the corresponding operators decay completely. The formalism presented in this section is very general. It can be applied to systems with arbitrary spectral functions. The only restriction is that the quantum system is coupled linearly to the environment.

As our formalism is quite different from other approaches to dissipative quantum systems, we discuss the exact solution for the dissipative harmonic oscillator in detail in Sect. 3. Since the Hamiltonian is quadratic, this problem can be solved for any spectral function of the bath. We show explicitly how the correlation functions are calculated. Although we only reproduce well–known results [17], the reader is encouraged to study this section. It provides the basic ideas for understanding how the information on the dissipative behaviour of the system can be contained in the continuous unitary transformation. Furthermore some of the results are needed later on.

In Sect. 4 we discuss the dissipative two–level system. We calculate the Fourier transform of the equilibrium correlation function. We are able to treat a super–Ohmic environment with arbitrary spectral properties or an Ohmic bath with small coupling. We obtain differential equations that can be solved numerically or discussed analytically for the qualitative behaviour. The numerical results are presented in Sect. 5. We compare our results with the non–interacting blip approximation. Both methods yield similar results for frequencies near the renormalized tunneling frequency. In addition we find the correct universal algebraic long–time behaviour of equilibrium correlation functions at zero temperature as put forward by Sassetti and Weiss [18]. Their generalized Shiba–relation [19] connecting intermediate and long time scales far beyond this "simple" universal long–time behaviour is also found to hold within numerical errors for the super–Ohmic bath.

Finally Sect. 6 contains some conclusions and an outline of possible future developments.

## 2 Flow equations: The general framework

In this section we develop a general method to treat model Hamiltonians for dissipative quantum systems. The method is an improvement and extension of the one used in our previous work on the spin–boson problem [14, 15]. The notation and the approximation we use are similar to our treatment of the Anderson single impurity model [16].

### 2.1 The model Hamiltonian

We study the properties of a small quantum system coupled to environmental degrees of freedom. As usual the environment is modeled by a set of non-interacting bosonic degrees of freedom, i.e. harmonic oscillators. The small quantum system is described by a general Hamiltonian $H_S$. It can for instance be the Hamiltonian of a particle in a confining potential or of a spin degree of freedom. In this section the quantum system is not specified. We assume that the small quantum system is coupled linearly to the bosonic degrees of freedom. The Hamiltonian of quantum system plus environment is given by

$$H = H_S + A \sum_k \lambda_k (b_k + b_k^\dagger) + \sum_k \omega_k : b_k^\dagger b_k : . \qquad (2.1)$$

The operators $b_k$ and $b_k^\dagger$ are usual bosonic creation and annihilation operators. In the sequel we will perform the limit where the number of bosonic degrees of freedom goes to infinity. The couplings $\lambda_k$ enter via the combination

$$J(\omega) = \sum_k \lambda_k^2 \delta(\omega - \omega_k). \qquad (2.2)$$

$J(\omega)$ is assumed to be a continuous function of $\omega$. A typical example is the so called Ohmic bath, where $J(\omega) = 2\alpha\omega g_c(\omega/\omega_c)$. $g_c(x)$ is a suitable cutoff function with $g_c(0) = 1$. Typical examples are $g_c(x) = \exp(-x)$ or $g_c(x) = \theta(1-x)$. The final results should not depend on the special choice of the cutoff function but be universal functions of some low–energy scale.

We treat the case of a general spectral function $J(\omega)$, but we will need some additional conditions on $J(\omega)$ to be introduced below. The operator $A$ has to be specified when one treats a special example. Concerning $H_S$ we assume in a large part of the paper that it has a pure point spectrum. Generalizations to the case where the spectrum of $H_S$ has continuous parts is possible if we assume that the ground state of $H_S$ is separated by a finite gap from the continuous part of the spectrum. Although we are able to treat the case of a free particle coupled to a bosonic bath exactly using our method, we exclude in the present paper the case where $H_S$ has a purely continuous spectrum. We do not discuss e.g. the case of a particle moving in a periodic potential.

The form of the Hamiltonian (2.1) contains the kind of problems discussed by Caldeira and Leggett [2]. There a particle in a confining potential and coupled to an environment was discussed. Caldeira and Leggett argued that as long as the moving particle perturbs each degree of freedom of the environment only weakly and the temperature is small, one can without loss of generality represent the environment by a set of harmonic oscillators. Furthermore they showed that apart from some pathological cases it is possible to choose the coupling between system and bath to be linear in the oscillator coordinates. Let us mention that recently Castro Neto and Caldeira [20] proposed other models for dissipative quantum systems that do not fall into the class of models described by (2.1).

## 2.2 Flow equations for the Hamiltonian

We now want to apply a continuous unitary transformation to the Hamiltonian (2.1). Such a transformation is defined by a generator $\eta$ that depends on a continuous variable $\ell$. The continuous unitary transformation is defined by (1.2) with the initial condition $H(0) = H$ in (2.1). The Hamiltonian and consequently the coupling constants become functions of $\ell$. For the generator of the continuous unitary transformation we make the following ansatz

$$\eta = i\sum_k A_k(b_k + b_k^\dagger) + \sum_k B_k(b_k - b_k^\dagger) + \sum_{k,q} \eta_{k,q} : (b_k + b_k^\dagger)(b_q - b_q^\dagger) : . \quad (2.3)$$

$A_k$ and $B_k$ are elements of the algebra spanned by $H_S(0)$ and $A$. In comparison to [14, 15], $\eta$ contains an additional term. This term is introduced to eliminate effective couplings between different bosonic modes. With this choice of $\eta$ one can construct the exact solution of the problem in the case of a harmonic oscillator or a free particle coupled to a bosonic bath, see Sect. 3. The commutator of $\eta$ with $H$ is given by

$$\begin{aligned}
[\eta, H] &= i\sum_k \omega_k A_k(b_k - b_k^\dagger) + \sum_k \omega_k B_k(b_k + b_k^\dagger) \\
&+ i\sum_k [A_k, H_S](b_k + b_k^\dagger) + \sum_k [B_k, H_S](b_k - b_k^\dagger) \\
&+ i\sum_{k,q} \lambda_q \langle [A_k, A] \rangle : (b_k + b_k^\dagger)(b_q + b_q^\dagger) : + i\sum_k \lambda_k [A_k, A](2n_k + 1) \\
&+ i\sum_{k,q} \lambda_q ([A_k, A] - \langle [A_k, A] \rangle) : (b_k + b_k^\dagger)(b_q + b_q^\dagger) : \\
&+ \sum_{k,q} \lambda_q [B_k, A] : (b_k - b_k^\dagger)(b_q + b_q^\dagger) : + \sum_k \lambda_k [B_k, A]_+ \\
&+ \sum_{k,q} \eta_{k,q} \omega_k : (b_k - b_k^\dagger)(b_q - b_q^\dagger) : + \sum_{k,q} \eta_{k,q} \omega_q : (b_k + b_k^\dagger)(b_q + b_q^\dagger) : \\
&+ 2\sum_{k,q} \eta_{k,q} \lambda_q (b_k + b_k^\dagger) A .
\end{aligned} \quad (2.4)$$

The expectation value $\langle . \rangle$ is defined by $\langle . \rangle = \frac{\text{Tr} . \exp(-\beta H_S)}{\text{Tr} \exp(-\beta H_S)}$. For zero temperature it is simply the ground state expectation value. We now choose $A_k$ and $B_k$ such that the coupling of the system to the bath has always the form given in (2.1) with an operator $A$ that does not depend on $\ell$. The $\ell$–dependence is contained in the couplings $\lambda_k$. This yields the following conditions for $A_k$ and $B_k$

$$A_k = \frac{i}{\omega_k}[B_k, H_S], \tag{2.5}$$

which eliminates couplings of system and environment in $[\eta, H]$ containing $b_k - b_k^\dagger$, and

$$\omega_k B_k + i[A_k, H_S] = -\lambda_k A f(\omega_k, \ell), \tag{2.6}$$

which guarantees that $A$ does not depend on $\ell$. Furthermore we choose $\eta_{k,q}$ so that couplings between different bath modes in (2.4) do not occur. To achieve this we need the following conditions

$$\eta_{k,q}\omega_k + \eta_{q,k}\omega_q = 0, \tag{2.7}$$

which eliminates couplings of the form $(b_k - b_k^\dagger)(b_q - b_q^\dagger)$, and

$$\eta_{k,q}\omega_q + \eta_{q,k}\omega_k + i\lambda_q \langle[A_k, A]\rangle + i\lambda_k \langle[A_q, A]\rangle = 0, \tag{2.8}$$

which eliminates couplings of the form $(b_k + b_k^\dagger)(b_q + b_q^\dagger)$. When these conditions are satisfied two terms in $[\eta, H]$ remain, namely the term in the fourth line and the first term in the fifth line of (2.4). They contain a coupling of the system to two bosonic modes. We neglect these terms. This approximation has to be justified a posteriori. The corresponding couplings are not present in the initial Hamiltonian. A rough estimate of these terms can be obtained if one integrates the coefficients over $\ell$.

A comparison of the terms in $[\eta, H]$ with the terms in the Hamiltonian yields the flow equations for the couplings $\lambda_k$ and for the Hamiltonian $H_S$ of the system

$$\frac{d\lambda_k}{d\ell} = -\lambda_k f(\omega_k, \ell) + 2\sum_q \eta_{k,q}\lambda_q, \tag{2.9}$$

$$\frac{dH_S}{d\ell} = i\sum_k \lambda_k[A_k, A](2n_k + 1) + \sum_k \lambda_k[B_k, A]_+. \tag{2.10}$$

The conditions on $A_k$, $B_k$ and on $\eta_{k,q}$ can be solved. We obtain

$$B_k = -\frac{1}{2i}\lambda_k f(\omega_k, \ell) \int_0^\infty d\tau e^{i\tau(\omega_k + i0_+)} \left(e^{i\tau H_S} A e^{-i\tau H_S} + e^{-i\tau H_S} A e^{i\tau H_S}\right), \tag{2.11}$$

$$A_k = -\frac{1}{2}\lambda_k f(\omega_k, \ell) \int_0^\infty d\tau e^{i\tau(\omega_k + i0_+)} \left(e^{i\tau H_S} A e^{-i\tau H_S} - e^{-i\tau H_S} A e^{i\tau H_S}\right), \tag{2.12}$$

and

$$\begin{aligned}
\eta_{k,q} &= i(\langle[A_k, A]\rangle \lambda_q + \langle[A_q, A]\rangle \lambda_k) \frac{\omega_q}{\omega_k^2 - \omega_q^2} \\
&= -\frac{i}{2}\frac{\lambda_k \lambda_q \omega_q}{\omega_k^2 - \omega_q^2} \left(f(\omega_k, \ell) \int_0^\infty d\tau e^{i\tau(\omega_k + i0_+)} \langle[e^{i\tau H_S} A e^{-i\tau H_S} - e^{-i\tau H_S} A e^{i\tau H_S}, A]\rangle \right. \\
&\quad \left. + f(\omega_q, \ell) \int_0^\infty d\tau e^{i\tau(\omega_q + i0_+)} \langle[e^{i\tau H_S} A e^{-i\tau H_S} - e^{-i\tau H_S} A e^{i\tau H_S}, A]\rangle \right).
\end{aligned} \tag{2.13}$$

With these choices of $A_k$ and $B_k$, the neglected terms in (2.4) are

$$-\frac{i}{2}\sum_{k,q}\lambda_q\lambda_k f(\omega_k,\ell):(b_k+b_k^\dagger)(b_q+b_q^\dagger):\int_0^\infty d\tau e^{i\tau(\omega_k+i0_+)}([e^{i\tau H_S}Ae^{-i\tau H_S}-e^{-i\tau H_S}Ae^{i\tau H_S},A]$$

$$-\langle[e^{i\tau H_S}Ae^{-i\tau H_S}-e^{-i\tau H_S}Ae^{i\tau H_S},A]\rangle) \tag{2.14}$$

$$+\frac{i}{2}\sum_{k,q}\lambda_q\lambda_k f(\omega_k,\ell):(b_k-b_k^\dagger)(b_q+b_q^\dagger):\int_0^\infty d\tau e^{i\tau(\omega_k+i0_+)}[e^{i\tau H_S}Ae^{-i\tau H_S}+e^{-i\tau H_S}Ae^{i\tau H_S},A]$$

Since $\langle[e^{i\tau H_S}Ae^{-i\tau H_S}+e^{-i\tau H_S}Ae^{i\tau H_S},A]\rangle=0$, the expectation value of these terms vanishes. Neglecting these terms yields a good approximation if the integral over $\ell$ of these terms is small. This is the case if $\lambda_k$ are small or if $\lambda_k$ fall off rapidly as functions of $\ell$. We will come back to this point at the end of this subsection.

The flow equations for $\lambda_k$ and $H_S$ can be rewritten using the $\ell$–dependent spectral function of the bath,

$$J(\omega)=\sum_k \lambda_k(\ell)^2\delta(\omega-\omega_k). \tag{2.15}$$

Notice $J(\omega,0)=J(\omega)$. Then the flow equations are

$$\frac{\partial J(\omega,\ell)}{\partial \ell} = -2J(\omega,\ell)f(\omega,\ell)-2iJ(\omega,\ell)\int d\omega'\frac{\omega' J(\omega',\ell)}{\omega^2-\omega'^2}$$

$$\left(f(\omega,\ell)\int_0^\infty d\tau e^{i\tau(\omega+i0_+)}\langle[e^{i\tau H_S}Ae^{-i\tau H_S}-e^{-i\tau H_S}Ae^{i\tau H_S},A]\rangle\right.$$

$$\left.+f(\omega',\ell)\int_0^\infty d\tau e^{i\tau(\omega'+i0_+)}\langle[e^{i\tau H_S}Ae^{-i\tau H_S}-e^{-i\tau H_S}Ae^{i\tau H_S},A]\rangle\right) \tag{2.16}$$

$$\frac{dH_S}{d\ell} = -\frac{i}{2}\int d\omega J(\omega,\ell)f(\omega,\ell)\int_0^\infty d\tau e^{i\tau(\omega+i0_+)}$$

$$((2n(\omega)+1)[e^{i\tau H_S}Ae^{-i\tau H_S}-e^{-i\tau H_S}Ae^{i\tau H_S},A]$$

$$-[e^{i\tau H_S}Ae^{-i\tau H_S}+e^{-i\tau H_S}Ae^{i\tau H_S},A]_+) \tag{2.17}$$

The integrand in the second term in (2.16) contains a singularity $\propto (\omega-\omega')^{-1}$. It occurs due to the corresponding singularity in $\eta_{k,q}$. The integral has to be interpreted as a principal value integral. The singularity in $\eta_{k,q}$ occurs because we tried to choose $\eta$ so that $H_S$ remains form invariant. This has the advantage that the number of flow equations is small. The price we have to pay is that the singularity can cause problems. We will come back to this point at the end of the next subsection.

Let us for a moment regard the trivial case where the commutator $[A,H_S]$ is initially a number $c$. Then $[A,H_S]=c$ for all $\ell$. Furthermore the neglected terms in $[\eta,H]$ vanish and the flow equation becomes exact. Integrating we obtain

$$H_{S\infty}-H_S(0)=-A^2\int d\omega \frac{J(\omega,0)}{\omega}. \tag{2.18}$$

In order to obtain a finite result we need the condition

$$\int d\omega \frac{J(\omega,0)}{\omega}<\infty. \tag{2.19}$$

This means that

$$J(\omega,0) \propto \omega^s, \quad s > 0 \quad \text{for small } \omega. \tag{2.20}$$

Furthermore $J(\omega,0)$ should contain a high–energy cutoff $\omega_c$. These conditions on $J(\omega,0)$ are usually taken for granted. They can e.g. be motivated from a usual adiabatic renormalization scheme, see Appendix 1. The case $s = 1$ is the Ohmic bath $J(\omega,0) = 2\alpha\omega g_c(\omega/\omega_c)$. $s > 1$ is called the super–Ohmic bath, $s < 1$ the sub–Ohmic bath.

A second point concerning (2.18) is that the term on the right–hand side is negative semi–definite. We have to introduce a corresponding counterterm in $H_S(0)$ since otherwise $H_{S\infty}$ can become unbounded from below. This is well–known from perturbative renormalization. A similar term occurs in the general situation as well. Writing $e^{i\tau H_S} A e^{-i\tau H_S}$ in (2.17) as a series of iterated commutators,

$$e^{i\tau H_S} A e^{-i\tau H_S} = A + i\tau[H_S, A] - \frac{1}{2}\tau^2[H_S, [H_S, A]] + ..., \tag{2.21}$$

the anticommutator in the last term in (2.17) yields a contribution $-A^2 \int d\omega \frac{J(\omega,0)}{\omega}$ to $H_{S\infty}$. One has to add a counterterm $A^2 \int d\omega \frac{J(\omega,0)}{\omega}$ to $H_S(0)$. This is well–known from adiabatic renormalization (compare Appendix 1).

## 2.3 Conditions on $f(\omega,\ell)$

The form of the function $f(\omega,\ell)$ has been left open until now. It is clear that $f(\omega,\ell)$ has to be chosen such that $J(\omega,\ell)$ tends to zero in the limit $\ell \to \infty$. In this subsection we analyse the flow equations (2.16, 2.17). The goal of this analysis is first to find an appropriate choice of $f(\omega,\ell)$ and second to see which conditions we need to be sure that the neglected terms in (2.14) are really small. We will show that $f(\omega,\ell)$ has to fulfil the conditions

$$f(\omega,\ell) \simeq \omega^2 \quad \text{for large } \omega, \tag{2.22}$$

$$f(\omega,\ell) \simeq (\omega - (\epsilon_n - \epsilon_m))^2 \quad \text{for } \omega \text{ near some resonance energy } (\epsilon_n - \epsilon_m). \tag{2.23}$$

The reader who is not interested in the details of the derivation of these conditions is encouraged to skip the remainder of this subsection.

Using (2.19) and the fact that $J(\omega,0)$ has to contain a high–energy cutoff $\omega_c$, we make for the moment an ansatz

$$J(\omega,\ell) = K\omega^s g_c(\omega/\tilde{\omega}(\ell)), \tag{2.24}$$

where $g_c(x)$ is some cutoff function with $g_c(0) = 1$. $\tilde{\omega}(\ell)$ is an $\ell$–dependent cutoff, $\tilde{\omega}(0) = \omega_c$. We assume that $\tilde{\omega}(\ell)$ decays monotonously with $\ell$. Such an ansatz is always possible for $\ell = 0$. It contains most interesting physical situations. If we neglect the second term in (2.16), such an ansatz is possible for $\ell > 0$ using

$$f(\omega,\ell) = \frac{g'(\omega/\tilde{\omega})}{2g(\omega/\tilde{\omega})} \frac{\omega}{\tilde{\omega}^2} \frac{d\tilde{\omega}}{d\ell}. \tag{2.25}$$

If we now use only the first two terms in the expansion (2.21) we obtain

$$\frac{dH_S}{d\ell} = \int d\omega K \frac{\omega^{s+1}}{\tilde{\omega}^2} g'(\omega/\tilde{\omega}) \frac{d\tilde{\omega}}{d\ell} \left( \frac{A^2}{\omega} + \frac{2n(\omega)+1}{2\omega^2} [[H_S, A], A] \right). \tag{2.26}$$

As long as $\tilde{\omega}$ is large compared to typical excitations of $H_S$ the solution of this equation is essentially equivalent to perturbative renormalization. Using (2.17) instead of (2.26) yields a result that is essentially equivalent to adiabatic renormalization. The main point is that as long as $\tilde{\omega}$ is large compared to typical excitations of $H_S$, couplings to bosonic modes with energies above $\tilde{\omega}$ are integrated out.

The question is now whether or not (2.24) can be used as a suitable approximation to (2.16). What is the effect of the second term in (2.16)? To study this question we use again the expansion (2.21) to write

$$i \int_0^\infty d\tau e^{i\tau(\omega+i0_+)} \langle [e^{i\tau H_S} A e^{-i\tau H_S} - e^{-i\tau H_S} A e^{i\tau H_S}, A] \rangle = \langle [[H_S, A], A] \rangle \frac{1}{\omega^2} + O(\frac{1}{\omega^3}). \tag{2.27}$$

Therefore we choose $f(\omega, \ell) \propto \omega^2$ for large $\omega$. This is possible using a cutoff function $g_c(x) = \exp(-x^2)$, which yields

$$f(\omega, \ell) = -\frac{\omega^2}{\tilde{\omega}^3} \frac{d\tilde{\omega}}{d\ell}. \tag{2.28}$$

We choose $\tilde{\omega} = (2\ell)^{-\frac{1}{2}}$ for $\ell > \omega_c^{-2}$, since then $f(\omega, \ell) = \omega^2$ does not depend on $\ell$ for $\ell > \omega_c^{-2}$. This is the first condition (2.22) on $f(\omega, \ell)$ mentioned above. The flow equation for $J(\omega, \ell)$ can now be written in the form

$$\frac{\partial J(\omega, \ell)}{d\ell} = 2 \frac{\omega^2}{\tilde{\omega}^2} \frac{d\tilde{\omega}}{d\ell} J(\omega, \ell) - 2 J(\omega, \ell) \frac{d\tilde{\omega}}{d\ell} \langle [[H_S, A], A] \rangle K \tilde{\omega}^{s-2} \int dx x^{\frac{s}{2}} \frac{e^{-x}}{x - \omega^2/\tilde{\omega}^2}. \tag{2.29}$$

In our ansatz (2.24) $K$ was assumed to be constant. Now, at least for $\omega^2/\tilde{\omega}^2 \ll 1$ this equation can be interpreted as an equation for the $\ell$-dependence of the coupling constant $K$,

$$\frac{dK}{d\ell} = -2 \langle [[H_S, A], A] \rangle \Gamma(\frac{s}{2}) \tilde{\omega}^{s-2} \frac{d\tilde{\omega}}{d\ell} K^2. \tag{2.30}$$

Since $\frac{d\tilde{\omega}}{d\ell} < 0$ and $\langle [[H_S, A], A] \rangle \leq 0$, $K$ decays monotonously. Integrating this equation yields

$$K(\ell) = \left( K(0)^{-1} + \frac{2\Gamma(\frac{s}{2}) \langle [[H_S, A], A] \rangle}{1 - s} (\tilde{\omega}(0)^{s-1} - \tilde{\omega}(\ell)^{s-1}) \right)^{-1} \quad \text{for } s \neq 1 \tag{2.31}$$

$$K(\ell) = \left( K(0)^{-1} - 2\Gamma(\frac{1}{2}) \langle [[H_S, A], A] \rangle \ln \frac{\tilde{\omega}(0)}{\tilde{\omega}(\ell)} \right)^{-1} \quad \text{for } s = 1. \tag{2.32}$$

Up to now we always used the expansion (2.21) to analyse the flow equations. This is possible for sufficiently large values of $\omega$. It is clear that the result (2.31, 2.32) together with the ansatz (2.24) applies only if $\omega$ lies not in the region of typical excitation energies of $H_S$. If this is the case the expansion (2.21) is no longer possible. In fact one should not interpret the $\ell$-dependence of $K$ as a coupling constant renormalization but as a correction to the $\ell$-dependence of $J(\omega, \ell)$.

To study the behaviour for smaller values of $\omega$, let us consider the case where $H_S$ has a pure point spectrum. If one calculates expressions $[e^{i\tau H_S} A e^{-i\tau H_S}]$ in the basis where $H_S$ is diagonal, the $\tau$-integrals in (2.16) and (2.17) can be evaluated. (2.16) can be written in the form

$$\frac{\partial J(\omega, \ell)}{\partial \ell} = -2 J(\omega, \ell) f(\omega, \ell) - 2i \sum_{n,m} a_{n,m} a_{m,n} \frac{\exp(-\beta \epsilon_n)}{Z_S} J(\omega, \ell) \int d\omega' \frac{\omega' J(\omega', \ell)}{\omega^2 - \omega'^2}$$
$$\times \left( f(\omega, \ell) \int_0^\infty d\tau e^{i\tau(\omega + i0_+)} (e^{i\tau(\epsilon_n - \epsilon_m)} - e^{-i\tau(\epsilon_n - \epsilon_m)}) \right.$$

$$+f(\omega', \ell) \int_0^\infty d\tau e^{i\tau(\omega'+i0_+)} \left(e^{i\tau(\epsilon_n-\epsilon_m)} - e^{-i\tau(\epsilon_n-\epsilon_m)}\right)\right)$$

$$= -2J(\omega,\ell)f(\omega,\ell) + 4\sum_{n,m} a_{n,m}a_{m,n}\frac{\exp(-\beta\epsilon_n)}{Z_S}J(\omega,\ell)\int d\omega' \frac{\omega' J(\omega',\ell)}{\omega^2-\omega'^2}$$

$$\left(\frac{f(\omega,\ell)}{\omega^2-(\epsilon_n-\epsilon_m)^2} + \frac{f(\omega',\ell)}{\omega'^2-(\epsilon_n-\epsilon_m)^2}\right) \quad (2.33)$$

Here $\epsilon_n$ are the eigenvalues of $H_S$, $Z_S = \sum_n \exp(-\beta\epsilon_n)$ and $a_{n,m}$ are the matrix elements of $A$ in the eigenbasis of $H_S$. This equation (and similarly the flow of $H_S$) contains divergencies $\propto (\omega-\epsilon_n+\epsilon_m)^{-1}$, where $\epsilon_n-\epsilon_m$ is an excitation energy of $H_S$. In order to avoid these divergencies, we demand that $f(\omega,\ell)$ behaves like $(\omega-\epsilon_n+\epsilon_m)^2$ for $\omega \approx \epsilon_n-\epsilon_m$. This is the second condition (2.23) mentioned above. In this way $f(\omega,\ell)$ becomes a function of the eigenvalues of $H_S$. If one neglects the second term in (2.16) a first consequence of this choice of $f(\omega,\ell)$ is that $J(\omega,\ell)$ does not change too much for small values of $\ell$ if $\omega \approx \epsilon_n-\epsilon_m$ whereas it decays if $\omega$ lies not in the vicinity of a resonance.

Now for values of $\omega$ large compared to typical excitation energies of $H_S$, the ansatz (2.24) still holds. But for values of $\omega$ small compared to typical excitation energies of $H_S$, the second term on the right–hand side of (2.33) is positive and the derivative of $J(\omega,\ell)$ becomes positive for not too large values of $\ell$. In terms of the couplings $\lambda_k$ this means that $\lambda_k$ increases for small values of $\omega_k$. Such a behaviour can cause problems since the neglected terms (2.14) in $[\eta,H]$ contain a factor $\lambda_k\lambda_q$. In general one can expect that the neglected terms are unimportant for indices $k$ and $q$ for which $\omega_k$ or $\omega_q$ are not too small. Whether or not they become important for small $\omega_k$ and $\omega_q$ depends on the behaviour of $J(\omega,0)$ for small $\omega$. A quantitative analysis in the case of the dissipative two–level system in Sect. 5 will show that no problems occur in the super–Ohmic case $J(\omega,0) \propto \omega^s$, $s > 1$, and in the Ohmic case $J(\omega,0) = 2\alpha\omega$ for sufficiently small $\alpha$.

The deeper reason for this infrared problem is our choice of $\eta$. Our goal was to construct $\eta$ in such a way that off–diagonal matrix elements in the Hamiltonian are eliminated fast if the difference of the corresponding diagonal matrix elements is large. Due to the denominator $(\omega_k-\omega_q)^{-1}$ in $\eta_{k,q}$ it happens that the derivative with respect to $\ell$ of some off–diagonal matrix elements becomes large even when the difference of the corresponding diagonal matrix elements is small.

To conclude our analysis of the general flow equations for the couplings and for the effective Hamiltonian (2.16) and (2.17), we have shown that for small values of $\ell$ the result of these equations is more or less equivalent to adiabatic renormalization. $(2\ell)^{-\frac{1}{2}}$ plays the role of the new high–energy cutoff. When $(2\ell)^{-\frac{1}{2}}$ is of the order of typical excitation energies of $H_S$, the couplings to higher energies in the total Hamiltonian $H(\ell)$ are already exponentially small. When $\ell$ becomes larger, $(2\ell)^{-\frac{1}{2}}$ can no longer be interpreted as a new high–energy cutoff: For such larger values of $\ell$ the flow equations yield an approximate diagonalization of the renormalized Hamiltonian.

In the next two sections the general method is applied to two special models, the dissipative harmonic oscillator and the dissipative two–level system (spin–boson model). Our method is exact for the dissipative harmonic oscillator. In the spin–boson model we restrict ourselves to the super–Ohmic case with arbitrary coupling and to the Ohmic case with small coupling for the reasons mentioned above.

## 2.4 Asymptotic behaviour

Let us now discuss the situation where $H_{S\infty}$ has a pure point spectrum. Let us first assume that all its eigenvalues are non–degenerate. We denote the eigenvalues by $\epsilon_{n\infty}$ and we assume $\epsilon_{n\infty} < \epsilon_{m\infty}$ for $n < m$. The first point we have to investigate is the asymptotic behaviour of the eigenvalues of $H_S(\ell)$. By $\epsilon_n$ we denote the eigenvalues of $H_S(\ell)$. As a first step we choose $f(\omega, \ell)$ such that

$$J(\omega, \ell) \propto J_{nm} \exp(-2C_{nm} \int_0^\ell (\omega - |\epsilon_m - \epsilon_n|)^2 d\ell') \tag{2.34}$$

for sufficiently large values of $\ell$. Such a choice of $f(\omega, \ell)$ is always possible. It is consistent with our discussion in the previous subsection, cf. (2.22, 2.23). Neglecting the second term in (2.16) we write (2.17) in the form

$$\begin{aligned}
\frac{dH_S}{d\ell} &= \frac{i}{4} \int d\omega \frac{\partial J(\omega, \ell)}{\partial \ell} \int_0^\infty d\tau e^{i\tau(\omega + i0_+)} \left( (2n(\omega) + 1)[e^{i\tau H_S} A e^{-i\tau H_S} - e^{-i\tau H_S} A e^{i\tau H_S}, A] \right. \\
&\quad \left. -[e^{i\tau H_S} A e^{-i\tau H_S} + e^{-i\tau H_S} A e^{i\tau H_S}, A]_+ \right).
\end{aligned} \tag{2.35}$$

The matrix elements of $A$ in the basis where $H_{S\infty}$ is diagonal are denoted as $a_{nm}$. The asymptotic behaviour of $\epsilon_n$ is now obtained from the equation

$$\begin{aligned}
\frac{d\epsilon_n}{d\ell} &\approx \frac{i}{2} \sum_m a_{nm} a_{mn} \int d\omega \frac{\partial J(\omega, \ell)}{\partial \ell} \int_0^\infty d\tau e^{i\tau(\omega + i0_+)} \\
&\quad \left( (2n(\omega) + 1)[e^{i\tau(\epsilon_n - \epsilon_m)} - e^{-i\tau(\epsilon_n - \epsilon_m)}] - [e^{i\tau(\epsilon_n - \epsilon_m)} + e^{-i\tau(\epsilon_n - \epsilon_m)}] \right) \\
&= \sum_m a_{nm} a_{mn} \int d\omega \frac{\partial J(\omega, \ell)}{\partial \ell} \frac{(2n(\omega) + 1)(\epsilon_n - \epsilon_m) + \omega}{\omega^2 - (\epsilon_n - \epsilon_m)^2}.
\end{aligned} \tag{2.36}$$

The main contribution to the integral over $\omega$ for large values of $\ell$ comes from values $\omega \approx |\epsilon_{n\infty} - \epsilon_{m\infty}|$. Therefore we obtain

$$\begin{aligned}
\frac{d\epsilon_n}{d\ell} &\propto -2 \sum_m a_{nm} a_{mn} J_{nm} C_{nm} \int d\omega \exp(-2C_{nm} \int_0^\ell (\omega - |\epsilon_n - \epsilon_m|)^2 d\ell') \\
&\quad \frac{\omega - |\epsilon_n - \epsilon_m|}{\omega + |\epsilon_n - \epsilon_m|} ((2n(\omega) + 1)(\epsilon_n - \epsilon_m) + \omega).
\end{aligned} \tag{2.37}$$

This equation is similar to the equation that determines the asymptotic behaviour for $\Delta$ in the spin–boson problem [14]. It can be analysed as in the paper by Lenz and Wegner [21]. We obtain $\epsilon_n = \epsilon_{n\infty} + \frac{c_n}{\sqrt{\ell}}$ and $J(\omega, \ell) \propto \frac{1}{\sqrt{\ell}}$ for $\omega = |\epsilon_{n\infty} - \epsilon_{m\infty}|$. It should be clear that the asymptotic behaviour can be different if $H_{S\infty}$ has some degeneracies.

## 2.5 Flow equations for observables

Our goal is to calculate dynamical correlation functions for dissipative quantum systems. As already mentioned in the introduction, the simple form of the final Hamiltonian $H(\ell = \infty)$ makes it easy to calculate expectation values. But it is clear that a given observable has to be transformed using the continuous unitary transformation. The transformation of an operator $O$ takes the form

$$\frac{dO}{d\ell} = [\eta, O] \tag{2.38}$$

In the generic case a solution of this equation will not be possible. We assume that initially $O(\ell = 0)$ is an observable of the quantum system and not of the bath. We use the following linear ansatz for $O(\ell)$

$$O = O_S + \sum_k O_{k+}(b_k + b_k^\dagger) + i\sum_k O_{k-}(b_k - b_k^\dagger) + ... \qquad (2.39)$$

We neglect higher normal–ordered terms in the expression for $O$. The ansatz is similar to our previous work on the spin–boson problem [14]. But the continuous unitary transformation is different. The flow equations for $O_S$ and $O_{k\pm}$ are

$$\frac{dO_S}{d\ell} = i\sum_k [A_k, O_{k+}](2n_k + 1) - i\sum_k [B_k, O_{k-}](2n_k + 1) + \sum_k [B_k, O_{k+}]_+ + \sum_k [A_k, O_{k-}]_+ \qquad (2.40)$$

and

$$\frac{dO_{k+}}{d\ell} = i[A_k, O_S] + 2\sum_q \eta_{k,q} O_{q+}, \qquad (2.41)$$

$$\frac{dO_{k-}}{d\ell} = -i[B_k, O_S] - 2\sum_q \eta_{q,k} O_{q-}. \qquad (2.42)$$

As mentioned above we assume $O_{k+}(0) = O_{k-}(0) = 0$. In contrast to Ref. [14] the operators $O_{k\pm}$ are coupled for different $k$. It is clear that it depends strongly on the initial operator $O(0)$ whether or not the ansatz (2.39) will give good results. We shall see in Sect. 3 that the ansatz becomes exact in the case of the dissipative harmonic oscillator if $O$ is e.g. the coordinate or the momentum of the particle. It is not exact if $O$ is e.g. a power of the coordinate. Using $O$ given in (2.39) we can now calculate equilibrium correlation functions of the form

$$C_{O^{(1)}O^{(2)}}(t) = \langle e^{itH}O^{(1)}e^{-itH}O^{(2)}\rangle = \operatorname{Tr}\left(\rho_{\mathrm{eq}} e^{itH}O^{(1)}e^{-itH}O^{(2)}\right) \qquad (2.43)$$

where $\rho_{\mathrm{eq}}$ is the equilibrium statistical operator. $O^{(r)}$ are observables of the system. It is useful to calculate the correlation function at $\ell = \infty$ since in this limit the system is decoupled from the bath. The time dependence and the equilibrium average can be calculated. The final result is

$$\begin{aligned}
C_{O^{(1)}O^{(2)}}(t) &= \langle e^{itH_{S\infty}} O^{(1)}_{S\infty} e^{-itH_{S\infty}} O^{(2)}_{S\infty}\rangle_{S\infty} \\
&+ \sum_k \langle e^{itH_{S\infty}} O^{(1)}_{k+\infty} e^{-itH_{S\infty}} O^{(2)}_{k+\infty} + e^{itH_{S\infty}} O^{(1)}_{k-\infty} e^{-itH_{S\infty}} O^{(2)}_{k-\infty}\rangle_{S\infty} \\
&\qquad (e^{-i\omega_k t}(n_k+1) + e^{i\omega_k t} n_k) \\
&+ i\sum_k \langle e^{itH_{S\infty}} O^{(1)}_{k+\infty} e^{-itH_{S\infty}} O^{(2)}_{k-\infty} + e^{itH_{S\infty}} O^{(1)}_{k-\infty} e^{-itH_{S\infty}} O^{(2)}_{k+\infty}\rangle_{S\infty} \\
&\qquad (e^{-i\omega_k t}(n_k+1) - e^{i\omega_k t} n_k). \qquad (2.44)
\end{aligned}$$

Let us consider this general expression for the correlation function at zero temperature. Let $O^{(i)}_{Snm}$ be the matrix elements of $O^{(i)}_{S\infty}$ and $O^{(i)}_{k\pm nm}$ the matrix elements of $O^{(i)}_{k\pm\infty}$ in the eigenbasis of $H_{S\infty}$. Then we obtain

$$\begin{aligned}
C_{O^{(1)}O^{(2)}}(t) &= \sum_n e^{it(\epsilon_{0\infty} - \epsilon_{n\infty})} O^{(1)}_{S0n} O^{(2)}_{Sn0} + \sum_k \sum_n e^{it(\epsilon_{0\infty} - \epsilon_{n\infty} - \omega_k)} \\
&\qquad (O^{(1)}_{k+0n} O^{(2)}_{k+n0} + O^{(1)}_{k-0n} O^{(2)}_{k-n0} + iO^{(1)}_{k+0n} O^{(2)}_{k-n0} + iO^{(1)}_{k-0n} O^{(2)}_{k+n0}). \qquad (2.45)
\end{aligned}$$

The Fourier transform $\hat{C}_{O^{(1)}O^{(2)}}(\omega)$ has the form

$$\hat{C}_{O^{(1)}O^{(2)}}(\omega) = \sum_n O^{(1)}_{S0n} O^{(2)}_{Sn0} \delta(\omega + \epsilon_{0\infty} - \epsilon_{n\infty}) + \sum_k \sum_n (O^{(1)}_{k+0n} O^{(2)}_{k+n0} + O^{(1)}_{k-0n} O^{(2)}_{k-n0}$$
$$+ i O^{(1)}_{k+0n} O^{(2)}_{k-n0} + i O^{(1)}_{k-0n} O^{(2)}_{k+n0}) \delta(\omega + \epsilon_{0\infty} - \epsilon_{n\infty} - \omega_k) \tag{2.46}$$

The first term in this expression yields an oscillating behaviour of $C_{O^{(1)}O^{(2)}}(t)$ for large $t$. In dissipative quantum systems this term is generically not present. Generically one has $O^{(1)}_{s0n} = 0$, $O^{(2)}_{sn0} = 0$ if $\epsilon_{n\infty} - \epsilon_{0\infty}$ lies in the support of $J(\omega, 0)$. In the examples in Sects. 3 and 4 we shall see that the asymptotic behaviour of $J(\omega, \ell)$ discussed above guarantees $O^{(1)}_{s0n} = 0$, $O^{(2)}_{sn0} = 0$. If $O_{s0n} \neq 0$, the derivative $\frac{dO_{k+}}{d\ell}$ in (2.41) would contain terms that vanish $\propto \ell^{-\frac{1}{4}}$ in the limit $\ell \to \infty$. Consequently $O_{k+}$ would diverge. This shows that for $\ell \to \infty$ the observable $O$ in (2.39) decays completely and one obtains $O_S(\ell \to \infty) = 0$. For small values of $\omega$ this yields

$$\hat{C}_{O^{(1)}O^{(2)}}(\omega) \approx O^{(1)}_{S00} O^{(2)}_{S00} \delta(\omega) \tag{2.47}$$
$$+ \sum_k (O^{(1)}_{k+00} O^{(2)}_{k+00} + O^{(1)}_{k-00} O^{(2)}_{k-00} + i O^{(1)}_{k+00} O^{(2)}_{k-00} + i O^{(1)}_{k-00} O^{(2)}_{k+00}) \delta(\omega - \omega_k).$$

For small values of $\omega_k$ the matrix elements of $A_k$ are $(A_k)_{nm} = i\lambda_k f(\omega_k, \ell) \tilde{a}_{nm}$. The $\tilde{a}_{nm}$ do not depend on $k$, they obey the condition $\sum_r \tilde{a}_{nr} H^S_{rm} - H^S_{nr} \tilde{a}_{rm} = a_{nm}$. The corresponding matrix elements of $B_k$ contain an additional factor $\omega_k$ and can be neglected. Therefore for small $\omega_k$, the $k$–dependence of $O_{k+\infty}$ is contained in a proportionality factor $\lambda_k(0)$. The matrix elements of $O_{k-\infty}$ contain an additional factor $\omega_k$ and can be neglected. Thus we obtain

$$\hat{C}_{O^{(1)}O^{(2)}}(\omega) - O^{(1)}_{S00} O^{(2)}_{S00} \delta(\omega) \propto J(\omega, 0) \tag{2.48}$$

for small values of $\omega$. Apart from possible oscillations given by the first terms in (2.46) the correlation function $C_{O^{(1)}O^{(2)}}(t)$ therefore shows generically an algebraic relaxation $\propto t^{-s-1}$ for long times. We will see that the generic behaviour of the correlation functions in the non–degenerate case is similar to the behaviour of correlation functions for the dissipative harmonic oscillator. This generic behaviour is obtained for operators $O_S$ that do not commute with $A_k$ for small $\omega_k$. The result (2.48) remains valid if we allow for some degeneracies in the spectrum of $H_S$. The main condition is that the ground–state of $H_S$ must be unique. The spectrum of $H_S$ can even contain a continuous part at higher energies. The general result (2.48) applies to the case of a particle in an arbitrary potential with bound states. The case of a periodic potential is not included.

## 3   The dissipative harmonic oscillator

The dissipative harmonic oscillator can be used as a good testing ground for any approximation scheme in dissipative quantum physics since it is exactly solvable. The reason is that the Hamiltonian is quadratic. A survey of the properties of the dissipative harmonic oscillator can be found in the book of Weiss [3]. The exact solution of this problem for a general spectral function $J(\omega, 0)$ was given by Haake et al. [17]. In this section we will show that our method is exact in the case of the dissipative harmonic oscillator. We are able to calculate all correlation functions. But our main reason for treating this model is to illustrate how our method works and how one is able to obtain a dissipative behaviour with a final Hamiltonian $H(\ell = \infty)$ that contains no coupling of the quantum system to the bath. Furthermore we will introduce methods to solve the flow equations that will be useful later in our treatment of the dissipative two–state system.

## 3.1 Flow equations for the Hamiltonian

Our general procedure is exact in this case since the neglected terms in $[\eta, H]$ in (2.4) vanish identically. If one chooses $\eta$ as above and $H_S = \frac{p^2}{2m} + \frac{m}{2}\Delta^2 q^2$, the term $\frac{p^2}{2m}$ is not changed, there is no mass renormalization in our approach. One only obtains a potential renormalization, which can be written as a renormalization of $\Delta$. Instead we write $H_S$ in the form

$$H = \Delta b^\dagger b + \sum_k \lambda_k (b + b^\dagger)(b_k + b_k^\dagger) + \sum_k \omega_k b_k^\dagger b_k + E_0 \tag{3.1}$$

and use a slightly different $\eta$ to preserve the form of the Hamiltonian.

$$\begin{aligned}\eta &= \sum_k \eta_k^{(1)}(b - b^\dagger)(b_k + b_k^\dagger) + \sum_k \eta_k^{(2)}(b + b^\dagger)(b_k - b_k^\dagger) \\ &+ \sum_{k,q} \eta_{k,q}(b_k + b_k^\dagger)(b_q - b_q^\dagger) + \eta_b(b^2 - b^{\dagger\,2})\end{aligned} \tag{3.2}$$

The commutator of $\eta$ with $H$ is easily calculated.

$$\begin{aligned}[\eta, H] &= \sum_k (\eta_k^{(1)}\Delta + \eta_k^{(2)}\omega_k + 2\eta_b\lambda_k + 2\sum_q \eta_{k,q}\lambda_q)(b + b^\dagger)(b_k + b_k^\dagger) \\ &+ \sum_k (\eta_k^{(1)}\omega_k + \eta_k^{(2)}\Delta)(b - b^\dagger)(b_k - b_k^\dagger) \\ &+ \sum_{k,q} (2\eta_k^{(1)}\lambda_q + \eta_{k,q}\omega_q)(b_k + b_k^\dagger)(b_q + b_q^\dagger) \\ &+ \sum_{k,q} \eta_{k,q}\omega_k(b_k - b_k^\dagger)(b_q - b_q^\dagger) \\ &+ 2\sum_k \eta_k^{(2)}\lambda_k(b + b^\dagger)(b + b^\dagger) + 2\eta_b\Delta(b^2 + b^{\dagger\,2})\end{aligned} \tag{3.3}$$

We choose

$$\eta_k^{(1)}\omega_k + \eta_k^{(2)}\Delta = 0 \tag{3.4}$$

to eliminate terms in $[\eta, H]$ containing $(b - b^\dagger)(b_k - b_k^\dagger)$,

$$\eta_{k,q}\omega_k + \eta_{q,k}\omega_q = 0 \tag{3.5}$$

to eliminate terms containing $(b_k - b_k^\dagger)(b_q - b_q^\dagger)$,

$$\eta_{k,q}\omega_q + \eta_{q,k}\omega_k + 2\eta_k^{(1)}\lambda_q + 2\eta_q^{(1)}\lambda_k = 0 \tag{3.6}$$

to eliminate terms containing $(b_k + b_k^\dagger)(b_q + b_q^\dagger)$, and

$$\eta_b \Delta + \sum_k \eta_k^{(2)}\lambda_k = 0 \tag{3.7}$$

to eliminate terms containing $(b^2 - b^{\dagger\,2})$. With these choices, the Hamiltonian remains form invariant. Comparing the coefficients on the left– and right–hand sides of $\frac{dH}{d\ell} = [\eta, H]$ and using $\lambda_k = O(1/\sqrt{N})$, we obtain the flow equations

$$\frac{d\Delta}{d\ell} = 4\sum_k \eta_k^{(2)}\lambda_k, \tag{3.8}$$

$$\frac{dE_0}{d\ell} = 2\sum_k \eta_k^{(2)}\lambda_k + 2\sum_k \eta_k^{(1)}\lambda_k, \tag{3.9}$$

$$\frac{d\omega_k}{d\ell} = O(\frac{1}{N}), \tag{3.10}$$

$$\frac{d\lambda_k}{d\ell} = \eta_k^{(1)}\Delta + \omega_k \eta_k^{(2)} + 2\sum_q \eta_{k,q}\lambda_q + 2\eta_b\lambda_k. \tag{3.11}$$

We choose

$$\eta_k^{(1)} = -\lambda_k \Delta \tilde{f}(\omega_k,\ell), \tag{3.12}$$

$$\eta_k^{(2)} = \lambda_k \omega_k \tilde{f}(\omega_k,\ell), \tag{3.13}$$

$$\eta_{k,q} = -\frac{\omega_q}{\omega_k}\eta_{q,k} = -\frac{2\lambda_k \lambda_q \Delta \omega_q}{\omega_k^2 - \omega_q^2}(\tilde{f}(\omega_k,\ell) + \tilde{f}(\omega_q,\ell)), \tag{3.14}$$

$$\eta_b = -\frac{1}{4\Delta}\frac{d\Delta}{d\ell}. \tag{3.15}$$

With

$$J(\omega,\ell) = \sum_k \lambda_k^2 \delta(\omega - \omega_k) \tag{3.16}$$

the flow equations can be written in the form

$$\frac{d\Delta}{d\ell} = 4\int d\omega\, \omega \tilde{f}(\omega,\ell) J(\omega,\ell), \tag{3.17}$$

$$\frac{dE_0}{d\ell} = 2\int d\omega (\omega - \Delta)\tilde{f}(\omega,\ell) J(\omega,\ell), \tag{3.18}$$

$$\begin{aligned}\frac{\partial J(\omega,\ell)}{\partial \ell} &= 2(\omega^2 - \Delta^2)\tilde{f}(\omega,\ell)J(\omega,\ell) \\ &\quad - 8\Delta J(\omega,\ell)\int d\omega'\omega'\frac{J(\omega',\ell)}{\omega^2 - \omega'^2}(\tilde{f}(\omega,\ell) + \tilde{f}(\omega',\ell)) \\ &\quad - \frac{J(\omega,\ell)}{\Delta}\frac{d\Delta}{d\ell}. \end{aligned} \tag{3.19}$$

In the preceeding section we discussed how $f(\omega,\ell) = -(\omega^2 - \Delta^2)\tilde{f}(\omega,\ell)$ has to be chosen. Since now the operator $A = b + b^\dagger$ induces only transitions between next–neighbour states of $H_S$, the only relevant excitation energy is $\Delta$. Following the arguments for the general case the suitable choice of $f(\omega,\ell)$ is $f(\omega,\ell) = (\omega - \Delta)^2$. We will not need to specify this special form of $f(\omega,\ell)$, but it is often helpful to have a special form of $f(\omega,\ell)$ in mind. We introduce

$$R(z,\ell) = \Delta \int d\omega \frac{\omega J(\omega,\ell)}{z - \omega^2} \tag{3.20}$$

and calculate its derivative with respect to $\ell$. This yields

$$\frac{\partial R(z,\ell)}{\partial \ell} = -\frac{1}{4}\frac{d\Delta^2}{d\ell} - (\Delta^2 + 4R(z,\ell) - z)\Delta \int d\omega \frac{\omega J(\omega,\ell)\tilde{f}(\omega,\ell)}{z - \omega^2}. \tag{3.21}$$

Obviously

$$\Delta^2 + 4R(\Delta_\infty^2, \ell) - \Delta_\infty^2 = 0 \quad (3.22)$$

is a solution if $R(z, \infty) = 0$. This yields a quadratic equation for $\Delta(\ell)$

$$\Delta(\ell)^2 = \Delta_\infty^2 - 4\Delta(\ell) \int d\omega\, \omega \frac{J(\omega, \ell)}{\Delta_\infty^2 - \omega^2}. \quad (3.23)$$

$\Delta_\infty$ can be determined from this equation for $\ell = 0$

$$\Delta_\infty^2 = \Delta_0^2 + 4\Delta_0 \int d\omega\, \omega \frac{J(\omega)}{\Delta_\infty^2 - \omega^2}. \quad (3.24)$$

We require for the reason of stability

$$\Delta_0 \geq 4 \int \frac{d\omega}{\omega} J(\omega, \ell). \quad (3.25)$$

This condition reflects the fact that the initial Hamiltonian has to contain a counter–term $A^2 \int \frac{d\omega}{\omega} J(\omega, \ell)$. For $\Delta_0 = 4 \int \frac{d\omega}{\omega} J(\omega, \ell)$ we obtain $\Delta_\infty = 0$. This is the case of a free particle. We discuss this special case at the end of this section. As an example that can be solved explicitly we let

$$J(\omega, 0) = \frac{\gamma^2 \omega \alpha}{\gamma^2 + \omega^2}. \quad (3.26)$$

If $J(\omega, 0)$ has this simple Drude–like behaviour we obtain

$$\Delta_\infty^2 = \Delta_0^2 - 2\pi \Delta_0 \gamma \alpha \frac{\gamma^2}{\gamma^2 + \Delta_\infty^2}. \quad (3.27)$$

The final solution is

$$\Delta_\infty^2 = \frac{\Delta_0^2 - \gamma^2}{2} + \sqrt{\frac{(\Delta_0^2 + \gamma^2)^2}{4} - 2\pi \Delta_0 \gamma^3 \alpha}. \quad (3.28)$$

## 3.2 Transformation of operators and correlation functions

Since $\eta$ is bilinear in the bosonic operators, the operators $O_{k\pm}$ in (2.39) are simply numbers if $O_S = b$ or $O_S = b^\dagger$. Therefore the neglected terms in (2.39) vanish identically. We make the ansatz

$$b(\ell) = \beta(\ell) b + \bar{\beta}(\ell) b^\dagger + \sum_k \alpha_k(\ell) b_k + \sum_k \bar{\alpha}_k(\ell) b_k^\dagger. \quad (3.29)$$

The flow equation for $b$ is

$$\frac{db}{d\ell} = [\eta, b]. \quad (3.30)$$

It leads directly to the flow equations for the parameters

$$\frac{d\beta}{d\ell} = 2\bar{\beta}\eta_b - \sum_k \eta_k^{(1)} \alpha_k + \sum_k \eta_k^{(2)} \alpha_k + \sum_k \eta_k^{(1)} \bar{\alpha}_k + \sum_k \eta_k^{(2)} \bar{\alpha}_k, \quad (3.31)$$

$$\frac{d\bar{\beta}}{d\ell} = 2\beta\eta_b + \sum_k \eta_k^{(1)} \alpha_k + \sum_k \eta_k^{(2)} \alpha_k - \sum_k \eta_k^{(1)} \bar{\alpha}_k + \sum_k \eta_k^{(2)} \bar{\alpha}_k, \quad (3.32)$$

$$\frac{d\alpha_k}{d\ell} = \eta_k^{(1)}\beta - \eta_k^{(2)}\beta + \eta_k^{(1)}\bar{\beta} + \eta_k^{(2)}\bar{\beta} - \sum_q \eta_{q,k}\alpha_q + \sum_q \eta_{k,q}\alpha_q + \sum_q \eta_{q,k}\bar{\alpha}_q + \sum_q \eta_{k,q}\bar{\alpha}_q, \quad (3.33)$$

$$\frac{d\bar{\alpha}_k}{d\ell} = \eta_k^{(1)}\beta + \eta_k^{(2)}\beta + \eta_k^{(1)}\bar{\beta} - \eta_k^{(2)}\bar{\beta} + \sum_q \eta_{q,k}\alpha_q + \sum_q \eta_{k,q}\alpha_q - \sum_q \eta_{q,k}\bar{\alpha}_q + \sum_q \eta_{k,q}\bar{\alpha}_q. \quad (3.34)$$

Introducing

$$r_\pm = \left(\frac{\Delta}{\Delta_0}\right)^{\pm\frac{1}{2}} (\beta \pm \bar{\beta}) \qquad (3.35)$$

$$s_{k,\pm} = \left(\frac{\omega_k}{\Delta_0}\right)^{\pm\frac{1}{2}} (\alpha_k \pm \bar{\alpha}_k) \qquad (3.36)$$

we obtain

$$\frac{dr_\pm}{d\ell} = 2\sum_k \left(\frac{\Delta}{\omega_k}\right)^{\frac{1}{2}} \eta_k^{(2)} s_{k,\pm} \qquad (3.37)$$

$$\frac{ds_{k,\pm}}{d\ell} = -2\left(\frac{\Delta}{\omega_k}\right)^{\frac{1}{2}} \eta_k^{(2)} r_\pm + 2\sum_q \left(\frac{\omega_k}{\omega_q}\right)^{\frac{1}{2}} \eta_{k,q} s_{q,\pm}. \qquad (3.38)$$

Since initially $r_+(0) = r_-(0) = 1$ and $s_{k,+}(0) = s_{k,-}(0) = 0$, we obtain $r_+ = r_-$ and $s_{k,+} = s_{k,-}$. In the following the subscripts $\pm$ are dropped. The quantity

$$r^2 + \sum_k s_k^2 = 1 \qquad (3.39)$$

is conserved. This conservation law reflects the fact that $[b(\ell), b^\dagger(\ell)] = 1$ for all $\ell$. In analogy to $R(z,\ell)$ we introduce the functions

$$S_1(z,\ell) = \sum_k \frac{s_k \lambda_k \sqrt{\Delta\omega_k}}{z - \omega_k^2}. \qquad (3.40)$$

$$S_2(z,\ell) = \sum_k \frac{s_k^2}{z - \omega_k^2}. \qquad (3.41)$$

Calculating the derivative with respect to $\ell$, we obtain

$$\frac{\partial S_2(z,\ell)}{\partial \ell} = -4(r + 2S_1(z,\ell)) \sum_k \frac{s_k \lambda_k \sqrt{\Delta\omega_k} \tilde{f}(\omega_k,\ell)}{z - \omega_k^2}, \qquad (3.42)$$

$$\frac{\partial S_1(z,\ell)}{\partial \ell} = -\frac{1}{2}\frac{dr}{d\ell} - 2(r + 2S_1(z,\ell)) \sum_k \frac{\lambda_k^2 \Delta\omega_k \tilde{f}(\omega_k,\ell)}{z - \omega_k^2}$$
$$-(\Delta^2 + 4R(z,\ell) - z) \sum_k \frac{s_k \lambda_k \sqrt{\Delta\omega_k} \tilde{f}(\omega_k,\ell)}{z - \omega_k^2}. \qquad (3.43)$$

This shows that

$$S_2(z,\ell) - \frac{(r + 2S_1(z,\ell))^2}{\Delta^2 + 4R(z,\ell) - z} \qquad (3.44)$$

does not depend on $\ell$. Since $\lambda_k(\infty) = 0$ one has $R(z, \infty) = 0$, $S_1(z, \infty) = 0$ and since $s_k(0) = 0$ one has $S_1(z, 0) = 0$, $S_2(z, 0) = 0$. Therefore we obtain

$$S_2(z, \infty) - \frac{r(\infty)^2}{\Delta_\infty^2 - z} = -(\Delta_0^2 + 4R(z, 0) - z)^{-1}. \tag{3.45}$$

This allows us to determine $r(\infty)^2$ and $s_k(\infty)^2$. An immediate consequence of (3.45) is that $r(\infty) = 0$ if $\Delta_\infty$ lies in the support of $J(\omega)$. This is the case we are interested in. We introduce the function

$$K(\omega) = \sum_k s_k(\infty)^2 \delta(\omega^2 - \omega_k^2), \tag{3.46}$$

which can be determined from $S_2(z, \infty)$ using

$$\begin{aligned} K(\omega) &= \frac{1}{\pi} \Im S_2(\omega^2 - i0^+, \infty) \\ &= -\frac{1}{\pi} \Im \left( \frac{1}{\Delta_0^2 - \omega^2 + i0^+ + 2\Delta_0 P \left( \int \frac{d\omega'^2 J(\omega', 0)}{\omega^2 - \omega'^2} \right) + 2\pi i \Delta_0 J(\omega, 0)} \right) \\ &= \frac{2\Delta_0 J(\omega, 0)}{\left( \Delta_0^2 - \omega^2 + 2\Delta_0 P \left( \int \frac{d\omega'^2 J(\omega', 0)}{\omega^2 - \omega'^2} \right) \right)^2 + 4\pi^2 \Delta_0^2 J^2(\omega, 0)}. \end{aligned} \tag{3.47}$$

$P(.)$ denotes the principal value of the integral. The first term in the denominator vanishes for $\omega = \Delta_\infty$. $K(\omega)$ shows a maximum for some $\omega < \Delta_\infty$. The shift of the maximum to a value below the resonance energy is due to damping. The behaviour of $K(\omega)$ for small $\omega$ is determined by the behaviour of $J(\omega)$ for small $\omega$. From (3.39) we obtain the sum rule

$$2 \int_0^\infty d\omega\, \omega K(\omega) = 1 \tag{3.48}$$

If $J(\omega, 0)$ shows a Drude–like behaviour (3.26) we obtain

$$K(\omega) = \frac{2\alpha\gamma^2 \Delta_0 \omega (\gamma^2 + \omega^2)}{\left( \Delta_0^2(\gamma^2 + \omega^2) - 2\pi\alpha\gamma^3 \Delta_0 - \omega^2(\gamma^2 + \omega^2) \right)^2 + 4\pi^2 \alpha^2 \Delta_0^2 \gamma^4 \omega^2} \tag{3.49}$$

$$= \frac{1}{\pi\gamma} \frac{(\Delta_0^2 - \Delta_\infty^2)(\gamma^2 + \Delta_\infty^2)\omega(\gamma^2 + \omega^2)}{((\Delta_\infty^2 - \omega^2)^2(\Delta_\infty^2 + \gamma^2 + \omega^2 - \Delta_0^2)^2 + \omega^2 \gamma^{-2}(\Delta_0^2 - \Delta_\infty^2)^2(\gamma^2 + \Delta_\infty^2)^2}. \tag{3.50}$$

Correlation functions can now be calculated. We define

$$C_{bb}(t) \stackrel{\text{def}}{=} \langle e^{iHt} b e^{-iHt} b \rangle_T = C_{b^\dagger b^\dagger} \stackrel{\text{def}}{=} \langle e^{iHt} b^\dagger e^{-iHt} b^\dagger \rangle_T, \tag{3.51}$$

$$C_{b^\dagger b}(t) \stackrel{\text{def}}{=} \langle e^{iHt} b^\dagger e^{-iHt} b \rangle_T, \quad C_{bb^\dagger} \stackrel{\text{def}}{=} \langle e^{iHt} b e^{-iHt} b^\dagger \rangle_T, \tag{3.52}$$

where $\langle . \rangle_T$ denotes the thermal equilibrium average. In the following we assume that $\Delta_\infty$ lies in the support of $J(\omega)$. Evaluating these correlation function at $\ell = \infty$ we obtain

$$C_{bb}(t) = \sum_k \alpha_k(\infty) \bar{\alpha}_k(\infty) (e^{-i\omega_k t} \langle b_k b_k^\dagger \rangle_T + e^{i\omega_k t} \langle b_k^\dagger b_k \rangle_T). \tag{3.53}$$

$\alpha_k(\infty)$ and $\bar\alpha_k(\infty)$ can be expressed using $s_k(\infty)$. The sum over $k$ can be written as an integral over $\omega$. This yields

$$C_{bb}(t) = \int d\omega \frac{\omega K(\omega)}{2} \left(\frac{\Delta_0}{\omega} - \frac{\omega}{\Delta_0}\right) \left(e^{-i\omega t} + \frac{2\cos\omega t}{e^{\omega/T} - 1}\right). \tag{3.54}$$

Similarly we obtain

$$C_{b^\dagger b}(t) = \int d\omega \frac{\omega K(\omega)}{2} \left[\left(\frac{\Delta_0}{\omega} + \frac{\omega}{\Delta_0}\right)\frac{2\cos\omega t}{e^{\omega/T} - 1} + \frac{4i\sin\omega t}{e^{\omega/T} - 1} + \left(\frac{\Delta_0}{\omega} + \frac{\omega}{\Delta_0} - 2\right)e^{-i\omega t}\right] \tag{3.55}$$

$$C_{bb^\dagger}(t) = \int d\omega \frac{\omega K(\omega)}{2} \left[\left(\frac{\Delta_0}{\omega} + \frac{\omega}{\Delta_0}\right)\frac{2\cos\omega t}{e^{\omega/T} - 1} - \frac{4i\sin\omega t}{e^{\omega/T} - 1} + \left(\frac{\Delta_0}{\omega} + \frac{\omega}{\Delta_0} + 2\right)e^{-i\omega t}\right]. \tag{3.56}$$

For $T = 0$ the long–time dynamics is determined by the low–frequency dependence of $J(\omega, 0)$. If we have $J(\omega, 0) \propto \omega^s$ for small $\omega$, then $K(\omega) \propto \omega^s$ for small $\omega$ as well and the correlation functions $C_{bb}(t)$, $C_{bb^\dagger}(t)$ and $C_{b^\dagger b}(t)$ behave as $t^{-s-1}$ for large $t$. The same behaviour is obtained for $C_{qq}(t) = \frac{1}{8m\Delta_0}(C_{b^\dagger b}(t) + C_{bb^\dagger}(t) + 2C_{bb}(t))$, whereas the velocity correlation function $C_{vv}(t) = \frac{\Delta_0}{8m}(2C_{bb}(t) - C_{b^\dagger b}(t) - C_{bb^\dagger}(t))$ decays as $t^{-s-3}$.

It is instructive to see how the effect of dissipation enters in the formulae. The main point is that in the denominator of $K(\omega)$ in the last expression in (3.47) the resonance $\omega = \Delta_\infty$ is damped by the second term. This term is due to the imaginary part of the denominator of $S_2(\omega^2 - i0^+, \infty)$ in (3.45), i.e. the imaginary part of $R(\omega^2 - i0^+, 0)$. The final Hamiltonian contains no information on the dissipative behaviour of the system. It is simply a sum of the Hamiltonian of the bath and the renormalized Hamiltonian of the system, which is a Hamiltonian of a harmonic oscillator with frequency $\Delta_\infty$. The information on the dissipative behaviour is completely contained in the continuous unitary transformation and consequently in the transformed operators $b(\infty)$ and $b^\dagger(\infty)$. These operators can be expressed as sums over bath operators alone, the corresponding term $O_S(\infty)$ in the general expression (2.39) vanishes since $r(\infty) = 0$.

Let us close this subsection with a remark concerning the spectrum of the Hamiltonian. It is well–known that the pure point spectrum of the original Hamiltonian $H(\ell = 0)$ contains only the ground state energy. The rest of the spectrum is absolutely continuous [22, 23]. In contrast, we obtain a final Hamiltonian $H(\ell = \infty)$ that has a pure point spectrum, which contains more eigenvalues, not only the ground state energy. At a first glance this seems to be a contradiction. But it is clear that for any finite value of $\ell$ the Hamiltonian $H(\ell)$ has only one eigenstate, namely the ground state. The argument by Arai [22] is sufficiently general and can be applied. The problem is that the excited states of $H(\ell = \infty)$ cannot be transformed back to $\ell = 0$ since these states are completely embedded into the continuum of bath states. Strictly speaking the limit $\ell \to \infty$ does not commute with the thermodynamic limit. This was already observed in the first paper on flow equation for Hamiltonians by Wegner [13]. If one wants to study the problem of the spectrum of $H$ rigorously, one has to stop the flow equations at some very large but finite value of $\ell$ and treat $J(\omega, \ell)$ as in [22]. The results for the correlation functions remain the same.

## 3.3 The free particle

The case of a free particle coupled to a dissipative environment has been studied e.g. by Aslangul et al. [24]. It is described by the general case if $H_{S\infty}$ is the Hamiltonian of a free particle, i.e. $H_{S\infty} = \frac{p^2}{2m}$. The treatment of this case is similar to the harmonic oscillator. The behaviour can be deduced from the calculations above by taking the limit $\Delta_\infty \to 0$. For the special example of a Drude–like behaviour (3.50) we obtain

$$K(\omega) = \frac{1}{\pi\gamma} \frac{\Delta_0^2 \gamma^2 \omega(\gamma^2 + \omega^2)}{\omega^4(\gamma^2 + \omega^2 - \Delta_0^2)^2 + \omega^2 \Delta_0^4 \gamma^2}. \tag{3.57}$$

Since now $K(\omega) \propto \omega^{-1}$, the correlation functions $C_{bb}$ etc. are not well–defined. We have to calculate the velocity correlation function

$$C_{vv}(t) = \langle v(t)v(0) \rangle_T. \tag{3.58}$$

It can be obtained for the harmonic oscillator as a suitable combination of the correlation functions for $b$ and $b^\dagger$, namely

$$C_{vv}(t) = \frac{\Delta_0}{8m}(2C_{bb}(t) - C_{b^\dagger b}(t) - C_{bb^\dagger}(t)). \tag{3.59}$$

A second point is that from (3.45) one obtains

$$r(\infty)^{-2} = 1 - 4\Delta_0 \int d\omega \frac{J(\omega, 0)}{\omega^3} \tag{3.60}$$

For $s \leq 2$ one has $r(\infty) = 0$, whereas $r(\infty) \neq 0$ for $s > 2$. Therefore $C_{vv}(t)$ contains an additional contribution

$$\frac{r(\infty)^2}{m^2}\langle p^2 \rangle = \frac{r(\infty)^2}{m} k_B T \tag{3.61}$$

This yields

$$C_{vv}(t) = \frac{1}{2m} \int d\omega \omega^2 K(\omega) \left( \frac{2\cos(\omega t)}{e^{\omega/T} - 1} + e^{-i\omega t} \right) + \frac{r(\infty)^2}{m} k_B T. \tag{3.62}$$

For the above example (3.26) we obtained $K(\omega) \propto \omega^{-1}$, which yields $C_{vv}(t) \propto t^{-2}$ for $T = 0$. Consequently one finds $C_{qq}(t) \propto \ln t$ for $t \to \infty$.

(3.62) can be used for any general $J(\omega, 0)$. For $\Delta_\infty = 0$ one has $\Delta_0 = 4\int d\omega \frac{J(\omega,0)}{\omega}$ and the general expression for $K(\omega)$ yields

$$K(\omega) = \frac{1}{\Delta_0} \frac{2J(\omega, 0)}{\omega^4 \left(4\int \frac{d\omega' J(\omega',0)}{\omega'(\omega^2 - \omega'^2)} - \frac{1}{\Delta_0}\right)^2 + 4\pi^2 \Delta_0^2 J^2(\omega, 0)}. \tag{3.63}$$

Thus if $J(\omega, 0) \propto \omega^s$ for small $\omega$, one has $K(\omega) \propto \omega^{-s}$ if $s \leq 2$ and $K(\omega) \propto \omega^{s-4}$ if $s \geq 2$. The velocity correlation function $C_{vv}(t)$ behaves like $t^{-3+s}$ for $s < 2$ and like $t^{1-s}$ for $s > 2$ and $T = 0$. For finite temperature one finds $C_{vv}(t) \propto t^{-2+s}$ for $s < 2$ and $C_{vv}(t) \to \frac{r(\infty)^2}{m} k_B T$ for $s > 2$. The factor $mr(\infty)^{-2}$ is often called a renormalized mass. Strictly speaking, however, neither in our approach nor in adiabatic renormalization there is mass renormalization.

## 4 The dissipative two–level system

The Hamiltonian of the dissipative two–level system is

$$H = -\frac{\Delta_0}{2}\sigma_x + \frac{1}{2}\sigma_z \sum_k \lambda_k (b_k^\dagger + b_k) + \sum_k \omega_k b_k^\dagger b_k + E_0. \tag{4.1}$$

This is of the general form (2.1) with

$$H_S = -\frac{\Delta}{2}\sigma_x + E_0, \quad A = \frac{1}{2}\sigma_z. \tag{4.2}$$

For a general discussion of various physical applications of this Hamiltonian we refer to the review of Leggett et al. [5] and to the book of Weiss [3]. Our main interest in this section is to calculate the time–dependent equilibrium auto–correlation function of $\sigma_z$ for this model. In [14, 15] we already studied this model using flow equations, but we were only able to calculate the correlation function for low frequencies. Using our improved transformation we can now discuss the correlation function in the entire frequency range.

## 4.1 Transformation of the Hamiltonian

Due to the simplicity of $H_S$ we are able to perform the transformation introduced for the general case in Sect. 2 without any additional approximations. Due to the transformation $\Delta$ and $E_0$ become functions of $\ell$. The integrals in the general expressions for $A_k$ and $B_k$ can be calculated. The result is

$$A_k = -\frac{1}{2}\lambda_k f(\omega_k,\ell)\frac{\Delta}{\omega_k^2 - \Delta^2}\sigma_y \tag{4.3}$$

and

$$B_k = -\frac{1}{2}\lambda_k f(\omega_k,\ell)\frac{\omega_k}{\omega_k^2 - \Delta^2}\sigma_z. \tag{4.4}$$

Using these expressions we can derive $\eta_{k,q}$

$$\eta_{k,q} = \frac{1}{2}\frac{\lambda_k \lambda_q \Delta \omega_q}{\omega_k^2 - \omega_q^2}\tanh\frac{\beta\Delta}{2}\left(\frac{f(\omega_k,\ell)}{\omega_k^2 - \Delta^2} + \frac{f(\omega_q,\ell)}{\omega_q^2 - \Delta^2}\right) \tag{4.5}$$

The flow equations for the spectral function of the bosonic bath and for the effective Hamiltonian of the quantum system are

$$\begin{aligned}\frac{\partial J(\omega,\ell)}{\partial \ell} &= -2f(\omega,\ell)J(\omega,\ell) \\ &\quad + 2\Delta\tanh\frac{\beta\Delta}{2}J(\omega,\ell)\int d\omega' \frac{\omega' J(\omega',\ell)}{\omega^2 - \omega'^2}\left(\frac{f(\omega,\ell)}{\omega^2 - \Delta^2} + \frac{f(\omega',\ell)}{\omega'^2 - \Delta^2}\right),\end{aligned} \tag{4.6}$$

$$\frac{dH_S}{d\ell} = \frac{1}{2}\int d\omega J(\omega,\ell)f(\omega,\ell)\left((2n(\omega)+1)\frac{\Delta}{\omega^2 - \Delta^2}\sigma_x - \frac{\omega}{\omega^2 - \Delta^2}\right). \tag{4.7}$$

This yields the flow equations for $\Delta$ and $E_0$.

$$\frac{d\Delta}{d\ell} = -\Delta \int d\omega J(\omega,\ell)(2n(\omega)+1)\frac{f(\omega,\ell)}{\omega^2 - \Delta^2}, \tag{4.8}$$

$$\frac{dE_0}{d\ell} = -\frac{1}{2}\int d\omega\,\omega J(\omega,\ell)\frac{f(\omega,\ell)}{\omega^2 - \Delta^2}. \tag{4.9}$$

The second term in the flow equation for $J(\omega,\ell)$ has been neglected in [14, 15]. The effect of this term has been discussed qualitatively in Sect. 2. Let us now investigate what happens in the special case of the dissipative two–level system. The flow equations for the dissipative two–level system depend on the temperature. The reason is that we neglect higher normal ordered interactions. We restrict our discussion to $T \ll \Delta_r$. Since $\Delta$ is the only excitation energy of $H_S$ in the case of a two–level system, the suitable choice for $f(\omega, \ell)$ following the general discussion in Sect. 2.3 is

$$f(\omega, \ell) = (\omega - \Delta)^2. \tag{4.10}$$

With this choice the equations for $\Delta$ and $E_0$ are free of divergencies. The only divergency is given by the denominator $(\omega - \omega')^{-1}$ in the integrand in the second term in (4.6). If $\omega$ is small compared to $\Delta$, the second term is positive. For small $\ell$ typical values of $\omega'$ are of the order $\ell^{-\frac{1}{2}}$. Therefore the second term is of the order $\Delta \ell^{-\frac{1}{2}} J(\omega, \ell)$, whereas the first term is of the order $\Delta^2 J(\omega, \ell)$. Thus the derivative of $J(\omega, \ell)$ is positive for small $\omega$ and not too large $\ell$. Usually $J(\omega, \ell)$ has a maximum for some $\omega$ of the order of $\Delta$. But due to the fact that $J(\omega, \ell)$ increases with $\ell$ for small $\omega$, $J(\omega, \ell)$ can be very asymmetric with respect to its maximum. As a consequence, the derivative of $\Delta$ in (4.8) can change its sign. On the other hand the structure of the neglected terms (2.14) is similar to the integrand in the general flow equation of $H_S$ (2.17). When the derivative of $\Delta$ changes its sign, the additional terms in $[\eta, H]$ neglected so far are no longer small. The numerical solution of the flow equations in Sect. 5 shows that our method is applicable in the super–Ohmic case $J(\omega, 0) \propto \omega^s$ with $s > 1$ for all coupling strengths, and for the Ohmic case $J(\omega, 0) = 2\alpha\omega$ with $\alpha \lessapprox 0.25$. In the following analysis we restrict ourselves to these cases.

Let us assume $T = 0$ for the moment. Before we calculate the correlation function $C(t)$ of the dissipative two–level system, we analyse the asymptotic behaviour of $\Delta$ and $J(\omega, \ell)$ for large $\ell$. If we neglect the second term in (4.6), we obtain exactly the same equations as in [14]. The asymptotic behaviour of this equations is given by $\Delta(\ell) - \Delta_\infty \propto \ell^{-\frac{1}{2}}$, $J(\omega, \ell) \propto \ell^{-\frac{1}{2}} \exp(-2(\omega - \Delta_\infty)^2 \ell)$ [14, 21]. The second term does not alter the asymptotic behaviour of $\Delta$ as long as the above conditions ($s > 1$ or $\alpha \lessapprox 0.25$ for the Ohmic case $s = 1$) are satisfied. As a consequence, the approximation

$$\ln \frac{\Delta_\infty}{\Delta(\ell)} \approx -\frac{1}{2} \int d\omega \frac{J(\omega, \ell)}{\omega^2 - \Delta_\infty^2} \tag{4.11}$$

is a good approximation to the flow equations (4.8) and (4.6) and can be used for $\ell = 0$ as a self–consistency condition for $\Delta_\infty$.

## 4.2 The dynamical correlation function

The main difference to our former approach [14, 15] becomes apparent in the flow equations for observables. For $\sigma_z(\ell)$ we make the same ansatz as before,

$$\sigma_z(\ell) = h(\ell)\sigma_z + \sigma_x \sum_k \chi_k(\ell)(b_k + b_k^\dagger) \,. \tag{4.12}$$

With this ansatz the flow equations for $h$ and $\chi_k$ are

$$\frac{dh}{d\ell} = -\Delta \sum_k \lambda_k \chi_k (2n_k + 1) \frac{f(\omega_k, \ell)}{\omega_k^2 - \Delta^2} \tag{4.13}$$

$$\frac{d\chi_k}{d\ell} = \Delta h \lambda_k \frac{f(\omega_k, \ell)}{\omega_k^2 - \Delta^2} + \sum_q \chi_q \frac{\lambda_k \lambda_q \Delta \omega_q}{\omega_k^2 - \omega_q^2} \tanh \frac{\beta \Delta}{2} \left( \frac{f(\omega_k, \ell)}{\omega_k^2 - \Delta^2} + \frac{f(\omega_q, \ell)}{\omega_q^2 - \Delta^2} \right). \tag{4.14}$$

The second term in (4.14) was not taken into account in [14, 15]. When this term is not present, $\chi_k(\ell)$ diverges if $\omega_k = \Delta_\infty$. The divergence is only logarithmic, but it is clearly unphysical. This was the main reason why we were not able to obtain reasonable results for correlation functions in a frequency range around $\Delta_\infty$ in Refs. [14, 15]. Due to the second term this divergence is now smeared out. This is the lesson from the dissipative harmonic oscillator, compare Sect. 3. As a consequence the Fourier transform of the correlation function

$$
\begin{aligned}
C(t) &\stackrel{\text{def}}{=} \frac{1}{2}\langle \sigma_z(t)\sigma_z(0) + \sigma_z(0)\sigma_z(t)\rangle_T \\
&= h^2(\infty)\cos(\Delta_\infty t) + \sum_k \chi_k^2(\infty)(2n_k+1)\cos(\omega_k t)
\end{aligned}
\tag{4.15}
$$

is well–defined.

Again due to normal–ordering, the flow equations for $\sigma_z$ depend on the temperature. The neglected terms in (4.12) are normal–ordered and do not contribute if we calculate the simple average $\langle \sigma_z \rangle$. But in the correlation function (4.15) we have to calculate an average of a product of two normal–ordered operators. Therefore the neglected terms in (4.12) may be important. This happens when the temperature becomes of the order of $\Delta_\infty$.

Let us now calculate the correlation function for low temperature $T \ll \Delta_r$. If one neglects the second term in (4.14), the quantity $h^2 + \sum_k \chi_k^2$ is conserved. Due to the second term in (4.14) this sum rule is only asymptotically fulfilled. We obtain

$$
\frac{d}{d\ell}(h^2 + \sum_k \chi_k^2) = 2\Delta \sum_{k,q} \frac{\chi_k \lambda_k \chi_q \lambda_q}{\omega_k + \omega_q} \left( \frac{f(\omega_k,\ell)}{\omega_k^2 - \Delta^2} + \frac{f(\omega_q,\ell)}{\omega_q^2 - \Delta^2} \right) \tag{4.16}
$$

Using the asymptotic behaviour of $\lambda_k$ one shows that the right–hand side falls of like $\ell^{-2}$, which shows that $h^2 + \sum_k \chi_k^2$ is asymptotically constant. As a consequence we obtain $h \to 0$ if $\Delta_\infty$ lies in the support of $J(\omega, 0)$. Using (4.13) one shows that $h(\ell)$ falls of like $\ell^{-\frac{1}{4}}$. Thus the asymptotic behaviour of $h$ is not affected by the second term on the right–hand side of (4.14). This term is only important for $\omega_k \approx \Delta$ and $\ell \approx \Delta_\infty^{-2}$. The fact that $h^2 + \sum_k \chi_k^2$ is asymptotically constant can be explained if one compares the flow equations for $h$ and $\chi_k$ with the flow equations for $r$ and $s_k$, (3.37) and (3.38). The latter can be written explicitly in the form

$$
\frac{dr}{d\ell} = -2 \sum_k \lambda_k s_k \frac{f(\omega_k,\ell)}{\omega_k^2 - \Delta^2} \sqrt{\Delta \omega_k} \tag{4.17}
$$

$$
\frac{ds_k}{d\ell} = 2r\lambda_k \frac{f(\omega_k,\ell)}{\omega_k^2 - \Delta^2}\sqrt{\Delta \omega_k} + 4 \sum_q s_q \frac{\lambda_k \lambda_q \Delta \sqrt{\omega_k \omega_q}}{\omega_k^2 - \omega_q^2} \left( \frac{f(\omega_k,\ell)}{\omega_k^2 - \Delta^2} + \frac{f(\omega_q,\ell)}{\omega_q^2 - \Delta^2} \right). \tag{4.18}
$$

$\lambda_k$ in the case of the dissipative two–level system corresponds to $2\lambda_k$ in the case of the dissipative harmonic oscillator due to the different definition of the couplings in the Hamiltonian. Furthermore $h$ corresponds to $r$ and $\chi_k$ corresponds to $s_k$. The difference between (4.13) and (4.17) is that the factor $\sqrt{\Delta \omega_k}$ in (4.17) is replaced by $\Delta$ in (4.13). Similarly $\sqrt{\Delta \omega_k}$ in the first term in (4.18) is replaced by $\Delta$ in (4.14) and $\sqrt{\omega_k \omega_q}$ in the second term is replaced by $\omega_q$. Since for large $\ell$ the couplings $\lambda_k$ differ only significantly from zero if $\omega_k$ lies near $\Delta$, the difference between the two sets of equations becomes small for large $\ell$. Therefore to obtain an expression for the Fourier transform of the correlation function $C(t)$, we proceed as in the case of the harmonic oscillator. We introduce the functions

$$
S_2(z,\ell) = \sum_k \frac{\chi_k^2}{z - \omega_k^2}, \tag{4.19}
$$

$$S_1(z,\ell) = \sum_k \frac{\sqrt{\omega_k \Delta} \chi_k \lambda_k}{z - \omega_k^2}, \tag{4.20}$$

$$S_0(z,\ell) = \sum_k \frac{\omega_k \lambda_k^2}{z - \omega_k^2}. \tag{4.21}$$

The conserved quantity (3.44) in the case of the dissipative harmonic oscillator corresponds to

$$S_2(z,\ell) - \frac{(h + S_1(z,\ell))^2}{\Delta^2 - z + \Delta S_0(z,\ell)} \tag{4.22}$$

for the dissipative two–level system. It is clear that this quantity is not conserved. But since for large $\ell$ the flow equations for the harmonic oscillator and for the dissipative two–level system are similar, this quantity will be approximately constant. Using the asymptotic behaviour of $\lambda_k$ one shows that the derivative of (4.22) behaves like $\ell^{-2}$. The asymptotic behaviour of the relevant quantities, $\chi_k$ and $h$ is much slower.

The physical reason for this approximate equivalence of the dissipative harmonic oscillator and the dissipative two–level system for large $\ell$ is the following simple observation: From (3.55) one obtains the mean occupation number for the harmonic oscillator at zero temperature

$$\langle b^\dagger b \rangle = \int d\omega \frac{\omega K(\omega)}{2} \left( \frac{\Delta_0}{\omega} + \frac{\omega}{\Delta_0} - 2 \right). \tag{4.23}$$

Let us consider a spectral function $J(\omega)$ that is strongly peaked for $\omega = \Delta_0$. Then $K(\omega)$ is strongly peaked as well and the integral (4.23) becomes small. *Since for large $\ell$ the spectral function $J(\omega, \ell)$ is strongly peaked, the dissipative harmonic oscillator behaves for large $\ell$ like a two–level system: The higher states of the harmonic oscillator are not occupied.* It is clear that this argument only holds for low temperatures $T \ll \Delta_r$.

Since $(h + S_1) \to 0$ for $\ell \to \infty$, we obtain

$$S_2(z,\infty) \approx S_2(z,\ell_0) - \frac{(h(\ell_0) + S_1(z,\ell_0))^2}{\Delta^2(\ell_0) - z + \Delta(\ell_0) S_0(z,\ell_0)} \tag{4.24}$$

for a sufficiently large value $\ell_0$. We let $\ell_0 = (2\lambda \Delta_\infty)^{-2}$. The right–hand side of (4.24) can be obtained numerically by integrating the flow equations, which will be done in Sect. 5. The one–sided Fourier transform

$$C(\omega) = \sum_k \chi_k^2(\infty) \coth(\frac{\beta \omega_k}{2}) \delta(\omega - \omega_k) \tag{4.25}$$

of the correlation function

$$C(t) = \int_0^\infty d\omega C(\omega) \exp(i\omega t) \tag{4.26}$$

can be obtained from $S_2(z,\infty)$ using $C(\omega) = -\frac{2\omega}{\pi} \Im S_2(\omega^2 - i0_+, \infty)$. Notice our unusual normalization condition as compared to the literature following from the initial condition $C(t=0) = 1$

$$\int_0^\infty d\omega \, C(\omega) \stackrel{!}{=} 1. \tag{4.27}$$

The analytical solution of the set of differential equations leading to $C(\omega)$ is not possible. A qualitative impression can be obtained by solving the linearized differential equations for $\lambda_k$ and $\chi_k$ up to some value $\ell_0 = (2\lambda\Delta_\infty)^{-2}$ with $\lambda = 0.2\ldots 0.5$. For smaller values of $\lambda$ the nonlinearities are too important and for larger values the asymptotic conserved quantity (4.22) cannot be used. In order to solve the differential equations (4.6) and (4.14) one stills has to make the additional approximation to replace $\Delta(\ell)$ and $h(\ell)$ on the right–hand side by $(1+\lambda)\Delta_\infty$ and $h = 1$, resp. This is reasonably good by comparison with the numerical solution. Putting everything together one finds the following result for the correlation function

$$C(\omega) \propto J(\omega,0)(1+\lambda)\Delta_\infty \left( \frac{(1+\lambda)\Delta_\infty(1-\exp(-(\omega-(1+\lambda)\Delta_\infty)^2\ell_0))^2}{(\omega^2-(1+\lambda)^2\Delta_\infty^2)^2} \right.$$

$$\left. + \frac{\omega\exp(-2(\omega-(1+\lambda)\Delta_\infty)^2\ell_0)}{((1+\lambda)^2\Delta_\infty^2-\omega^2+(1+\lambda)\Delta_\infty\Re S_0(\omega^2-i0+,\ell_0))^2 + \frac{1}{4}(1+\lambda)^2\Delta_\infty^2 J(\omega,\ell_0)^2} \right) \quad (4.28)$$

Only the two most important terms from (4.24) are used here, the first coming from the solution of the linearized differential equations and the second term coming from the conserved quantity proportional to $h(\ell_0)^2$. Eq. (4.28) reflects the qualitative features of the numerical solution for $\omega \to 0$, $\omega \approx \Delta_r$ and $\omega \gg \Delta_r$. Due to all the necessary approximations it is not a very good approximation to the numerical solution beyond such qualitative features that are independent of the arbitrary parameter $\lambda$. We have therefore made no attempt to improve this analytical result by e.g. taking more terms into account.

## 5 Numerical results

In order to obtain quantitative results for the correlation function $C(\omega)$, we have to integrate the flow equations numerically. The strategy is obvious: First of all we numerically integrate the differential equations up to some sufficiently large value $\ell_0$. This gives us $S_2(z,\ell_0)$ in (4.24). Then we add the asymptotic conserved quantity to obtain $S_2(z,\infty)$ and finally $C(\omega)$.

How this is done explicitly will be explained in the next subsection. In particular it has to be established that the final result is independent of $\ell_0$. Subsection 5.2 contains quantitative results for the correlation function for Ohmic and super–Ohmic baths. All calculations are done for zero temperature but can easily be extended to $T < \Delta_\infty$. Such effects of nonzero but small temperature will be briefly explained in the text. The unrenormalized tunneling frequency $\Delta_0$ always defines the energy scale with $\Delta_0 = 1$.

### 5.1 Technicalities and tests

We solve the flow equations for the imaginary parts of the functions $S_0(z,\ell)$, $S_1(z,\ell)$ and $S_2(z,\ell)$ defined in the following manner

$$J_i(\omega,\ell) \stackrel{\text{def}}{=} -\frac{2}{\pi}\Im S_i(\omega^2-i0_+,\ell), \qquad i = 0,1,2. \quad (5.1)$$

For convenience we write $J_0(\omega,\ell) \stackrel{\text{def}}{=} J(\omega,\ell)$ in this section. From $J_i(\omega,\ell)$ we can also reconstruct the real part of $S_i(\omega,\ell)$

$$\Re S_i(\omega^2-i0_+,\ell) = P\int d\omega' \frac{\omega' J_i(\omega',\ell)}{\omega^2-\omega'^2}. \quad (5.2)$$

The complete set of differential equations that has to be solved numerically for the spin–boson problem is

$$\frac{d\Delta}{d\ell} = -\Delta \int d\omega \coth\frac{\beta\omega}{2} J_0(\omega,\ell)\frac{\omega-\Delta}{\omega+\Delta} \quad (5.3)$$

$$\frac{\partial J_0(\omega,\ell)}{\partial \ell} = -2(\omega-\Delta)^2 J_0(\omega,\ell) + 2\Delta \tanh\frac{\beta\Delta}{2} J_0(\omega,\ell) \int d\omega' J_0(\omega',\ell) I(\omega,\omega',\ell) \quad (5.4)$$

$$\frac{dh}{d\ell} = -\int d\omega \sqrt{\Delta\omega} \coth\frac{\beta\omega}{2} J_1(\omega,\ell) \frac{\omega-\Delta}{\omega+\Delta} \quad (5.5)$$

$$\frac{\partial J_1(\omega,\ell)}{\partial \ell} = -(\omega-\Delta)^2 J_1(\omega,\ell) + \Delta h\sqrt{\frac{\Delta}{\omega}} J_0(\omega,\ell)\frac{\omega-\Delta}{\omega+\Delta} + \frac{1}{2\Delta}\frac{d\Delta}{d\ell} J_1(\omega,\ell) \quad (5.6)$$

$$+\Delta \tanh\frac{\beta\Delta}{2}\Big(J_1(\omega,\ell)\int d\omega' J_0(\omega',\ell) I(\omega,\omega',\ell)$$

$$+J_0(\omega,\ell)\int d\omega' \sqrt{\frac{\omega'}{\omega}} J_1(\omega',\ell) I(\omega,\omega',\ell)\Big)$$

where

$$I(\omega,\omega',\ell) = \frac{\omega'}{\omega^2-\omega'^2}\left(\frac{\omega-\Delta(\ell)}{\omega+\Delta(\ell)} + \frac{\omega'-\Delta(\ell)}{\omega'+\Delta(\ell)}\right). \quad (5.7)$$

$J_2(\omega,\ell)$ can be obtained via the relation

$$J_2(\omega,\ell) = \frac{1}{\Delta}\frac{J_1^2(\omega,\ell)}{J_0(\omega,\ell)}. \quad (5.8)$$

After integrating these equations up to $\ell_0$, we add the approximate conserved quantity from (4.24) and obtain $C(\omega)$.

Now the differential equations for $J_0(\omega,\ell)$ and $J_1(\omega,\ell)$ have to be discretized for certain values $\omega_i$ and can only then be solved numerically. Typically we have used about 200 bath modes. Obviously it makes no sense to use equidistant values of $\omega_i$, rather one has to sample the vicinity of $\Delta_\infty$ much denser: For large $\ell$ the curves $J_0(\omega,\ell)$ and $J_1(\omega,\ell)$ are only nonvanishing close to $\Delta_\infty$. This is reminiscent of numerical renormalization, however, there the vicinity of $\omega=0$ is sampled denser although the main spectral weight might be somewhere else (compare e.g. Ref. [11]). In all cases we had at least a minimum of 10 bath modes within the half width of the final spectral function $J_0(\omega,\ell_0)$.

The quality of the numerical routines and this discretization procedure can be tested conveniently for the dissipative harmonic oscillator discussed in Sect. 3. The set of differential flow equations is very similar to (5.3)–(5.6), also the conserved quantity has the same structure. We have compared the numerical result for the function $K(\omega)$ with the exact solution from (3.47) for various values of $\ell_0 = (2\lambda\Delta_\infty)^{-2}$. The agreement always turned out to be excellent. A typical result for a Drude–like spectral function (3.26) can e.g. be seen in Fig. 1. The maximum deviation of the curves is less than 2% for all values of $\omega$.

For the dissipative harmonic oscillator the conserved quantity is an exactly conserved quantity as we have seen in Sect. 3. This situation was different for the spin–boson problem where it was only an asymptotic conserved quantity. Hence a second important test for the consistency of our approach is to check that the final result $C(\omega)$ in the spin–boson model is independent of $\ell_0 = (2\lambda\Delta_\infty)^{-2}$ if $\lambda$ is sufficiently small.

We have e.g. investigated this for the Ohmic bath with the parameters chosen as in Figs. 2a–c. In every diagram the curve $C(\omega,\ell_0)$ is the correlation function without using the conserved quantity and $C(\omega)$ is the final result after adding the approximate conserved quantity. One sees that for smaller $\lambda$ more and more spectral weight is contained in the numerical solution and less in the conserved quantity. However, the final results $C(\omega)$ for the different values of $\lambda$ agree very well as can be seen in Fig. 2d. The spectral function $J_0(\omega,\ell_0) \stackrel{\text{def}}{=} J(\omega,\ell_0)$ becomes more and more peaked around $\Delta_\infty$ for smaller $\lambda$ as can also be seen in Fig. 2a–c. Hence the physical reasoning underlying the approximate conserved quantity (the harmonic oscillator being equivalent to a spin–boson model if $J_0(\omega,\ell_0)$ is strongly peaked around $\Delta_\infty$, compare the discussion following (4.23)) is confirmed very well. Similar tests have been made for the other sets of parameters discussed in subsection 5.2 as well always leading to the same conclusion. As a suitable compromise between computing time and the fact that we only have an asymptotic conserved quantity we have used $\lambda = 0.1$ for all quantitative results in subsection 5.2.

A final test for the consistency of our approximations is provided by the normalization condition $\int_0^\infty C(\omega)\,d\omega \stackrel{!}{=} 1$ derived from the initial condition $C(t=0) = 1$. Remember that according to (4.16) this is not exactly fulfilled. However, for all curves in Fig. 2 and also in the next subsection the sum rule is fulfilled with an error of less than 1%. This agreement can be considered excellent and also gives support to the ansatz for the transformed observable $\sigma_z$ according to (4.12) as very little spectral weight is lost.

## 5.2 Quantitative results

*Ohmic bath:*

Fig. 3 contains the correlation function $C(\omega)$ for an Ohmic bath $J(\omega) = 2\alpha\omega\Theta(\omega_c - \omega)$ with various values of $\alpha$. $C(\omega)$ is plotted as a function of $\omega/\Delta_r$ with the low–energy scale $\Delta_r$. $\Delta_r$ is here used as a fit parameter to identify the positions of the various peaks (therefore the areas under the curves are necessarily different). The half width of the curves gives the dimensionless $Q$–factor of the damped oscillations. For $\alpha = 0$ one has only a $\delta$–peak corresponding to undamped oscillations and in Fig. 3 one observes how this peak decays as a function of increasing coupling $\alpha$.

Unfortunately, for larger values of $\alpha$ than contained in Fig. 3 we did no longer observe that the final result for $C(\omega)$ was independent of $\ell_0$. This problem can be traced back to the small energy denominators in the generator $\eta_{k,q}$ in (4.5). We have already discussed this possible problem in Sect. 2.3 and we refer to this section for more information. Hence using the present approximations our approach is limited to small couplings $\alpha \lessapprox 0.25$.

Now let us analyse our results for such small couplings in more detail. In Fig. 4 we compare the flow equation result with the correlation function obtained by the NIBA–approximation (see e.g. Ref. [5]). The quantitative agreement for intermediate time scales is very good. Differences are apparent in the long–time behaviour, i.e. the behaviour for small $\omega$. This can be more clearly seen in the response function $S(\omega) = C(\omega)/(\pi\omega)$ in Fig. 4: The NIBA–result diverges for $\omega \to 0$ which is an indication of the wrong long–time behaviour $C(t) \propto t^{-2(1-\alpha)}$, whereas $S(0)$ is finite in the flow equation framework. As expected the flow equations therefore yield the correct universal long–time behaviour $C(t) \propto t^{-2}$. Hence we agree with the NIBA where the NIBA can probably be trusted and we disagree where the NIBA is known to be wrong.

A very sensitive test is provided by the Shiba–relation [19] generalized to the spin–boson model in Ref. [18]. For an Ohmic bath this reads

$$\lim_{\omega \to 0} \frac{C(\omega)}{\omega} \stackrel{!}{=} 2\alpha\,(2\chi_0)^2. \tag{5.9}$$

The static susceptibility $\chi_0$ is extracted with a Kramer's–Kronig relation and a fluctuation–dissipation theorem from the correlation function $C(\omega)$

$$\chi_0 = \frac{1}{2}\int_0^\infty \frac{C(\omega)}{\omega}d\omega. \tag{5.10}$$

Notice that our normalization (4.27) $\int_0^\infty d\omega\, C(\omega) = 1$ differs from the definition in Ref. [18] which makes up for a factor $\pi$ in (5.9). In Table 1 relation (5.9) is tested for various parameters and small $\alpha$.

| $\alpha$ | $\omega_c$ | $\lim_{\omega\to 0} C(\omega)/\omega$ | $2\alpha\,(2\chi_0)^2$ | %error |
|---|---|---|---|---|
| 0.01  | 40.0  | 0.0209 | 0.0212 | 2%  |
| 0.01  | 100.0 | 0.0213 | 0.0222 | 2%  |
| 0.025 | 40.0  | 0.0644 | 0.0625 | 3%  |
| 0.025 | 100.0 | 0.0682 | 0.0660 | 3%  |
| 0.05  | 40.0  | 0.18   | 0.16   | 10% |
| 0.05  | 100.0 | 0.20   | 0.18   | 10% |
| 0.1   | 40.0  | 0.94   | 0.69   | 25% |
| 0.1   | 100.0 | 0.94   | 0.69   | 25% |

Table 1: Generalized Shiba–relation for an Ohmic bath as found from the numerical solution of the flow equations ($\Delta_0 = 1$). The typical numerical error of both numbers obtained numerically is estimated as 3%.

One observes that for very small couplings the Shiba–relation is fulfilled within numerical errors. However, the deviations grow with increasing $\alpha$ which can again be traced back to the small energy denominator problem for the Ohmic bath in (4.5). The typical effect of this is that $\lim_{\omega\to 0} C(\omega)/\omega$ comes out too large. We leave this problem for future investigation and note that for the super–Ohmic bath investigated below the generalized Shiba–relation turns out to be fulfilled with very good accuracy in the whole parameter range. Since the error in the Ohmic Shiba–relation can be traced back to small frequencies which carry very little spectral weight in the correlation function, this error does have little implications for the accuracy of the correlation function (for the NIBA the error is in fact infinite, still the NIBA probably makes sense on intermediate time scales for small couplings).

The low–energy scale of the spin–boson model with Ohmic coupling is set by

$$\Delta_r = c\, \Delta_0 \left(\frac{\Delta_0}{\omega_c}\right)^{\frac{\alpha}{1-\alpha}}, \tag{5.11}$$

where $c$ is some constant of order 1 as has already been shown in our previous paper [14]. This theoretical prediction was found to be well confirmed from the numerical results, where for small $\alpha$ the energy scale $\Delta_r$ plays the role of a parameter determined from the maximum of $C(\omega)$. One expects to find universal correlation functions as a function of $\omega/\Delta_r$ for $\omega \ll \omega_c$. In Fig. 5 the response function $S(\omega)$ is plotted as a function of this rescaled energy. Universality is confirmed with excellent accuracy.

Let us mention that the curves in Fig. 5 can be compared directly with numerical renormalization results of Costi et al. [11]. Qualitatively the curves look similar for these parameters, however, the maximum in the NRG framework is considerably lower: $S(\Delta_r)/S(0) \approx 4.0$ from NRG whereas we find $S(\Delta_r)/S(0) \approx 6.0$ using flow equations. We believe this problem can be traced back to problems inherent to numerical renormalization approaches for small couplings $\alpha$: One has to resolve a sharp peak in the correlation function where due to logarithmic discretization very few bath modes lie. From Fig. 4 it is apparent that discretization is no problem in the flow equation framework.

A final remark about small temperatures $T < \Delta_\infty$. The main effect is that the correlation functions $C(\omega)$ in Fig. 3 have to be multiplied with a factor $\coth(\beta\omega/2)$, for a qualitative discussion the effect of nonzero temperature in Eqs. (5.3)–(5.6) is negligible. Therefore $\lim_{\omega \to 0} C(\omega) = \text{const.} \neq 0$ for nonzero temperature leading to the expected exponential long–time decay of $C(t)$.

*Super–Ohmic bath:*

For a super–Ohmic bath the coupling function can be parametrized as

$$J(\omega) = K^{1-s}\omega^s \Theta(\omega_c - \omega), \quad s > 1 \tag{5.12}$$

with a coupling constant $K$ with dimension energy. The low–energy scale $\Delta_r$ is set by

$$\Delta_r = c\,\Delta_0 \, \exp\left(-\frac{1}{2(s-1)}\left(\frac{\omega_c}{K}\right)^{s-1}\right) \tag{5.13}$$

where $c$ is a constant of order 1. For the numerical calculations we have used a super–Ohmic baths with $s = 2, 3$. Extension to other values of $s > 1$ is unproblematic and yields similar results.

In Fig. 6 the correlation function $C(\omega)$ is plotted as calculated from flow equations or the NIBA. The qualitative agreement is reasonable for intermediate time scales. Anyway it is problematic to justify the NIBA–approximation for super–Ohmic baths since the blips cannot be considered as a dilute gas [5]. As in the Ohmic case, an important difference shows up in the long–time behaviour. Since $\lim_{\omega \to 0} C(\omega) = \text{const.} \neq 0$ in the NIBA (compare Fig. 6) one obtains an exponential long–time decay of $C(t)$, whereas the flow equation result is $C(\omega) \propto \omega^s$ for small $\omega$, hence $C(t) \propto t^{-s-1}$ for long times.

Again a sensitive test for the long–time behaviour is provided by the generalized Shiba–relation also put forward for super–Ohmic baths in Ref. [18]. It reads

$$\lim_{\omega \to 0} \frac{C(\omega)}{\omega^s} \stackrel{!}{=} K^{1-s}(2\chi_0)^2. \tag{5.14}$$

In Table 2 this relation is tested for various parameters.

One observes that the generalized Shiba–relation holds within numerical errors.[1] Considering the fact that the Shiba–relation is a highly nontrivial universal property connecting intermediate and long time scales far beyond the "simple" universal $C(t) \propto t^{-1-s}$ behaviour, this result can be considered very satisfactory. This supports the conclusion that our ansatz for the flow of the Hamiltonian and the observables already contains all the "relevant" physics and is probably very close to the exact solution. Let us also remark that for super–Ohmic baths the Shiba–relation cannot be tested with numerical renormalization group methods.

Correlation functions for various parameters of a super–Ohmic bath are plotted in Fig. 7. For increasing coupling corresponding to smaller values of $K$, the damping becomes stronger and stronger resulting in an increasing $Q$–factor. (Of course no phase transition occurs for some critical coupling as it does in the Ohmic case for $\alpha_c = 1$.)

A final remark about nonzero but small temperature $T < \Delta_r$. Again the main modification is that the zero–temperature curves are multiplied with the factor $\coth(\beta\omega/2)$. However, for $s > 1$ this still leads to an algebraic long–time decay of $C(t)$ proportional to $t^{-s}$. That is for very small temperature one still has an algebraic decay for super–Ohmic baths and not an exponential one. This might be surprising at first sight, however, for the exactly solvable dissipative harmonic oscillator one makes the same observation.[2]

---

[1] Notice that the deviations are systematic. But that cannot be interpreted with certainty due to our present numerical error of the used computer routines.

| $s$ | $K$ | $\omega_c$ | $\lim_{\omega \to 0} \frac{C(\omega)}{\omega^s}$ | $K^{1-s}(2\chi_0)^2$ | %error |
|---|---|---|---|---|---|
| 2 | 40.0 | 40.0 | 0.0622 | 0.0628 | 1% |
| 2 | 20.0 | 40.0 | 0.317 | 0.314 | 1% |
| 2 | 10.0 | 40.0 | 3.96 | 4.03 | 2% |
| 2 | 5.0 | 40.0 | 268 | 269 | 0.1% |
| 2 | 120.0 | 80.0 | 0.0151 | 0.0157 | 4% |
| 2 | 80.0 | 80.0 | 0.0314 | 0.0330 | 5% |
| 2 | 40.0 | 80.0 | 0.166 | 0.172 | 4% |
| 2 | 20.0 | 80.0 | 2.19 | 2.32 | 5% |
| 3 | 40.0 | 20.0 | 0.000653 | 0.000710 | 8% |
| 3 | 20.0 | 20.0 | 0.00380 | 0.00399 | 5% |
| 3 | 10.0 | 20.0 | 0.0594 | 0.0628 | 8% |

Table 2: Generalized Shiba–relation for a super–Ohmic bath as found from the numerical solution of the flow equations ($\Delta_0 = 1$). The typical numerical error of both numbers obtained numerically is estimated as 3%.

## 6   Conclusions

This has been a long and in parts probably tedious paper for the reader. However, the approach put forward here is quite complementary to common wisdom in dissipative quantum systems and required more explanations and justifications than usually necessary.

Let us sum up the main results of this paper. We have investigated dissipative quantum systems with the specific examples of the dissipative harmonic oscillator and the spin–boson model using infinitesimal unitary transformations. The Hamiltonian for such problems is of the form

$$H = H_S + H_B + H_{SB}, \tag{6.1}$$

where $H_S$ is the small system coupled via $H_{SB}$ to the environment described by the bath Hamiltonian $H_B$. With a continuous sequence of unitary transformations, a unitarily equivalent Hamiltonian $H_\infty$ has been found where system and bath are decoupled

$$H_\infty = H_{S\infty} + H_B. \tag{6.2}$$

In order to do this some approximations that neglect higher ("irrelevant") normal–ordered interactions generated by the unitary transformations have been used for the spin–boson model. The quality of these approximations seems to be quite good as we have seen in this paper (see also below). Notice that our approach is systematic in the sense that higher order terms can successively be taken into account.

---

[2] It is not necessarily true that the long–time behaviour for nonzero temperature is an exponential decay set by the smallest Matsubara frequency as sometimes claimed in the literature.

*What is perhaps most surprising at first sight is that such a program going from (6.1) to (6.2) using unitary transformations can be carried out at all.* The key question is where dissipation can enter in a description like (6.2). The answer lies hidden in the fact that the observables have to be transformed as well under the unitary transformation. If the discrete eigenstates of $H_{S\infty}$ are embedded in the continuum of bath states, we have found that generically these observables "decay" completely under the sequence of unitary transformations. That means the term describing the original observable with respect to (6.1) has vanished completely and been transformed into other terms with respect to the transformed Hamiltonian (6.2). These new terms show typical dissipative behaviour (e.g. decay of correlations in time) and in this manner dissipation enters into our results.

For the exactly solvable dissipative harmonic oscillator this programme could be carried through without any approximations. Within certain approximations the spin–boson model could also be mapped onto an unitarily equivalent spin–boson model by decoupling all except the nearly resonant bath modes. This procedure is depicted in Fig. 8. Large energy differences are decoupled for small flow parameters $\ell$ and smaller energy differences only later for larger flow parameters. This yields the fundamentally important separation of energy scales underlying e.g. renormalization theory. Notice the main difference from adiabatic renormalization where decoupling starts from the UV–cutoff and is never concerned with the low–lying states, also their energy difference to the discrete states might become comparable to the decoupled states at some stage of the renormalization procedure. Hence in order to obtain an effective spin–boson model with only nearly resonant couplings remaining, the flow equation technique is more suited. Using infinitesimal unitary transformations it is also apparent that the observables have to be transformed as well as is essential when states with an energy difference of order $\Delta_r$ are beginning to be decoupled.

As we have explained such a two–level model with only nearly resonant couplings becomes effectively equivalent to a dissipative harmonic oscillator as the higher states of the harmonic oscillator are not occupied. The exact solution of the dissipative harmonic oscillator could then be used as an asymptotic conserved quantity to calculate the correlation functions of the effective spin–boson model. This program could be carried out for an Ohmic bath with small coupling ($\alpha \lessapprox 1/4$) and for super–Ohmic baths with arbitrary coupling. The main restriction was that the temperature should be much smaller than the renormalized tunneling frequency $\Delta_r$ as only then the equivalence to the dissipative harmonic oscillator is possible.

We have found agreement of our results with the NIBA on intermediate time scales set by $\Delta_r$ where the NIBA can probably be trusted. The long–time behaviour of the spin–spin correlation function that is not accessible by the simple NIBA has an universal algebraic decay put forward by Sassetti and Weiss [18]. Their generalized Shiba–relation connecting intermediate and long–time scales was found to be fulfilled within numerical errors for super–Ohmic baths. Considering the fact that this generalized Shiba–relation is a highly nontrivial universal property far beyond a "simple" universal algebraic decay $C(t) \propto t^{-s-1}$, this indicates that the approximations in this paper retain all the "relevant" physics.

Compared to our approach the other methods mentioned in the Introduction have specific shortcomings in the low temperature regime. E.g. quantum Markov processes yield an entirely different long–time (low–energy) behaviour. There correlation functions typically show an exponential decay, even for the exactly solved harmonic oscillator. This can only be obtained from a Hamiltonian with a spectral function of the bath that is strongly modified with respect to its low energy properties as can easily be seen from the exact solution. However, a quantum Markov process can at least be derived from a Hamiltonian description of system plus bath since the complete positivity condition is satisfied [25]. For the NIBA already this point seems less clear.

At present it is unclear whether our method with some other approximations can also be used in the left–open parts of the parameter space of the spin–boson model. In particular it would be interesting to investigate temperatures $T > \Delta_r$.

**Acknowledgements**

A.M. thanks Volker Bach for many useful discussions during the special semester at the Erwin Schrödinger Institut in Vienna. S.K. thanks H.–P. Breuer for pointing out Ref. [6]. This work was partly supported by the Erwin Schrödinger Institut and the Deutsche Forschungsgemeinschaft.

# Appendix 1

In this appendix we briefly describe the adiabatic renormalization procedure applied to the model (2.1) in the spirit of Wilson [7]. In a renormalization procedure the Hamiltonian is divided into two parts $H = H_0 + H_1$, where $H_0$ describes the high–energy modes and $H_1$ describes the low–energy modes. The Hamiltonian is then mapped to an effective Hamiltonian $H^{(1)}$ that operates on the Hilbert space of the low–energy modes and has, at least approximately, the same low–energy spectrum as $H$. If necessary this procedure can be iterated and one constructs a series of Hamiltonians $H^{(n)}$. In the simplest perturbative renormalization scheme one sets $H^{(1)} = P_0 H_1 P_0$ where $P_0$ is the projector onto the ground states of $H_0$. This is the first step in a degenerate perturbational treatment, higher orders can be calculated similarly. In the adiabatic renormalization scheme, the Hamiltonian (2.1) is first transformed using a unitary transformation

$$U_0 = \exp(\sum_{\omega_k \in I_0} \frac{\lambda_k}{\omega_k} A(b_k - b_k^\dagger)) \tag{A.1}$$

where $I_0 = [\tilde{\omega}, \omega_c]$. The unitary transformation does not affect $A$, but only $H_S$. The transformed Hamiltonian becomes

$$H_\lambda^{(1)} = H_S^{(1)} + \sum_{\omega_k \in I_0} \omega_k b_k^\dagger b_k \tag{A.2}$$

where

$$H_S^{(1)} = U_0^\dagger H_S U_0 - A^2 \sum_{\omega_k \in I_0} \frac{\lambda_k^2}{\omega_k}. \tag{A.3}$$

This Hamiltonian is now projected onto the ground states of $\sum_{\omega_k \in I_0} \omega_k b_k^\dagger b_k$. Thereby we obtain an effective Hamiltonian which has the same form as the original Hamiltonian, but with a new cutoff $\tilde{\omega}$ and with a new renormalized $H_S$. $\tilde{\omega}$ has to be large compared to typical excitation energies of $H_S$. The projection onto the ground states of $\sum_{\omega_k \in I_0} \omega_k b_k^\dagger b_k$ corresponds to the first order of a degenerate perturbational treatment. Let $P$ be the projector onto the ground states of $\sum_{\omega_k \in I_0} \omega_k b_k^\dagger b_k$. The next term in the perturbational treatment is

$$P U_0^\dagger H_S U_0 (1-P) (\sum_{\omega_k \in I_0} \omega_k b_k^\dagger b_k)^{-1} (1-P) U_0^\dagger H_S U_0 P. \tag{A.4}$$

This term is of the order of a typical squared excitation energy of $H_S$ divided by $\tilde{\omega}$. It is small compared to the first term as long as $\tilde{\omega}$ is large compared to typical excitation energies of $H_S$.

This estimate shows one of the advantages of the adiabatic renormalization scheme: Corrections can be estimated and are small as long as the effective cutoff is large.

The procedure can now be iterated. We can introduce a new $\tilde{\omega}$ that is smaller than the old one but large compared to the typical excitation energies of the renormalized $H_S$. Alternatively, $\tilde{\omega}$ can be determined self–consistently to be large compared to typical excitation energies of the renormalized $H_S$. The final result of this procedure is

$$H_{\text{eff}} = H_{S,\text{eff}} + \sum_{k:\omega_k<\tilde{\omega}} \left(A\lambda_k(b_k + b_k^\dagger) + \omega_k b_k^\dagger b_k\right) \tag{A.5}$$

with

$$H_{S,\text{eff}} = \langle e^{-A\sum_{\tilde{\omega}<\omega_k<\omega_c} \frac{\lambda_k}{\omega_k}(b_k-b_k^\dagger)} H_S e^{A\sum_{\tilde{\omega}<\omega_k<\omega_c} \frac{\lambda_k}{\omega_k}(b_k-b_k^\dagger)} \rangle_B - A^2 \sum_{\tilde{\omega}<\omega_k<\omega_c} \frac{\lambda_k^2}{\omega_k}. \tag{A.6}$$

$\langle . \rangle_B$ denotes an average over the bath. This new Hamiltonian contains no high–energy scale. The initial Hamiltonian has to contain a counterterm of the form $A^2 \sum_k \lambda_k^2/\omega_k$. This is well–known in the theory of dissipative quantum systems. The main problem is that it is still difficult to calculate dynamical correlation functions using the renormalized Hamiltonian since it still has a complicated structure.

# References

[1] Caldeira, A.O. and Leggett, A.J.: Physica **121A**, 587 (1983); **130A**, 374(E) (1985)

[2] Caldeira, A.O. and Leggett, A.J.: Ann. Phys. (NY) **149**, 374 (1984); **153**, 445(E) (1984)

[3] Weiss, U.: *Quantum Dissipative Dynamics.* Series in Modern Condensed Matter Physics, Vol. 2, World Scientific, Singapore 1993

[4] Feynman, R.P. and Vernon, F.L.: Ann. Phys. (NY) **243**, 118 (1963)

[5] Leggett, A. J., Chakravarty, S., Dorsey, A. T., Fisher, M. P. A., Garg, A. and Zwerger, W.: Rev. Mod. Phys. **59**, 1 (1987); *erratum* Rev. Mod. Phys. **67**, 725 (1995)

[6] Alicki, R. and Lendi, K.: *Quantum Dynamical Semigroups and Applications.* Lecture Notes in Physics Vol. 286, Springer, Berlin 1987

[7] Wilson, K.G.: Phys Rev **140**, B445 (1965)

[8] Bray, A.J. and Moore, N.A.: Phys. Rev. Lett. **49**, 1545 (1982)

[9] Chakravarty, S.: Phys. Rev. Lett. **49**, 682 (1982)

[10] Wilson, K.G.: Rev. Mod. Phys. **47**, 773 (1975)

[11] Costi, T.A. and Kieffer, C.: Phys. Rev. Lett. **76**, 1683 (1996)

[12] Glazek, S.D. and Wilson, K.G.: Phys. Rev. D **49**, 4214 (1994)

[13] Wegner, F.: Ann. Physik (Leipzig) **3**, 77 (1994)

[14] Kehrein, S.K., Mielke, A. and Neu, P.: Z. Phys. B **99**, 269 (1996)

[15] Kehrein, S.K. and Mielke, A.: Phys. Lett. A **219**, 313 (1996)

[16] Kehrein, S.K. and Mielke, A.: 'Theory of the Anderson impurity model: The Schrieffer–Wolff transformation re–examined.' ESI–Preprint 270, cond–mat/9510145 (1995), to appear in Ann. Phys. (NY)

[17] Haake, F. and Reibold, R.: Phys. Rev. **A32**, 2462 (1985)

[18] Sassetti, M. and Weiss, U.: Phys. Rev. Lett. **65**, 2262 (1990)

[19] Shiba, H. Prog. Theor. Phys. **54**, 967 (1975)

[20] Castro Neto, A.H. and Caldeira, A.O.: Phys. Rev. Lett. **67**, 1960 (1991)

[21] Lenz, P. and Wegner, F.: 'Flow equations for the electron–phonon interaction.' Preprint cond–mat/9604087 (1996)

[22] Arai, J. Math. Phys. **22**, 2539-2548 and 2549-2552 (1981)

[23] Bach, V., Fröhlich, J. and Sigal, I.M.: Lett. Math. Phys. **34**, 183 (1995)

[24] Aslangul, C., Pottier, N., Saint-James, D.: J. Stat. Phys. **40**, 167 (1985)

[25] Davies, E.B.: *Quantum Theory of Open Systems.* Academic Press, London 1976.

# Figure captions

Fig. 1. Testing the quality of the numerical routines for the exactly solvable dissipative harmonic oscillator. Here the spectral function is Drude–like $J(\omega) = \gamma^2 \omega \alpha/(\gamma^2 + \omega^2)$ with parameters $\alpha = 0.1, \gamma = 1.0$ and $\Delta_0 = 1.0$. The function $K(\omega)$ defined in (3.46) determines all the equilibrium correlation functions and is plotted from the exact solution (3.50) and the numerical solution of the flow equations. Here $\lambda = 0.1$ was chosen, but no dependence on $\lambda$ can be observed within numerical errors.

Fig. 2. Results from the numerical solution of the flow equations for the spin–boson model with an Ohmic spectral function $J(\omega) = 2\alpha\omega\Theta(\omega_c - \omega)$. Parameters are $\alpha = 0.1, \omega_c = 10.0$ and $\Delta_0 = 1.0$. In Fig. 2a the flow equations are numerically integrated up to $\lambda = 0.2$ and then the asymptotic conserved quantity is added to $C(\omega, \ell_0)$. In Fig. 2b $\lambda = 0.1$ is chosen and in Fig. 2c we have integrated even further until $\lambda = 0.05$. The final result for the correlation function $C(\omega)$ is within small errors independent of these values $\lambda$ as can be seen in Fig. 2d.

Fig. 3. Correlation functions $C(\omega)$ for the spin–boson model with an Ohmic bath ($\omega_c = 10.0, \Delta_0 = 1.0$) and various values of $\alpha$. The functions $C(\omega)$ are plotted as functions of $\omega/\Delta_r$ with the low–energy scale $\Delta_r$ used as a fit parameter to identify the peaks. $\Delta_r$ is set by (5.11).

Fig. 4. Comparison of the flow equation result for the correlation function $C(\omega)$ with the NIBA–curve. An Ohmic bath is used with parameters $\omega_c = 10.0, \alpha = 0.1$ and $\Delta_0 = 1.0$. For intermediate time scales the NIBA–results agree with the flow equation results, however, the long–time behaviour of the NIBA is wrong as can be seen from the divergence of the response function $S(\omega)$ for $\omega \to 0$. The data points for $C(\omega)$ show how the bath energies are discretized in the flow equation framework.

Fig. 5. The normalized response function $S(\omega)$ is plotted as a function of the rescaled energy $\omega/\Delta_r$ where $\Delta_r$ is again used as a fit parameter to identify the peaks. An Ohmic bath is used with parameters $\alpha = 0.1, \Delta_0 = 1.0$ and various values of $\omega_c$. Universality is confirmed with excellent accuracy.

Fig. 6. Comparison of the flow equation result for the correlation function $C(\omega)$ with the NIBA–result. An super–Ohmic bath with a spectral function like in Eq. (5.12) is used. Parameters are $s = 2, K = 5, \Delta_0 = 1.0$ and $\omega_c = 10.0$. The main difference to the NIBA is again the long–time behaviour as can be clearly seen in the log–log–plot.

Fig. 7. Correlation functions $C(\omega)$ for the spin–boson model with a super–Ohmic bath of type (5.12). Parameters are $s = 2, \Delta_0 = 1.0, \omega_c = 10.0$ and various values of $K$. $\Delta_r$ is again used as a fit parameter to identify the peaks. It is set by the low–energy scale derived from adiabatic renormalization for $s = 2$: $\Delta_r = c\Delta_0 \exp(-\omega_c/(2K))$. $c$ is a constant of order 1.

Fig. 8. Sketch of the flow of the Hamiltonian and of a generic observable as a function of $\ell$ in different regimes. For small values of $\ell$ the system finds the low–energy scale. In this region the flow of the observables is negligible, and the flow equations are equivalent to adiabatic renormalization. In the other regions, the flow of the observable becomes important. The $\omega$–scale in the plots of $J(\omega, \ell)/J(\omega, 0)$ is logarithmic.

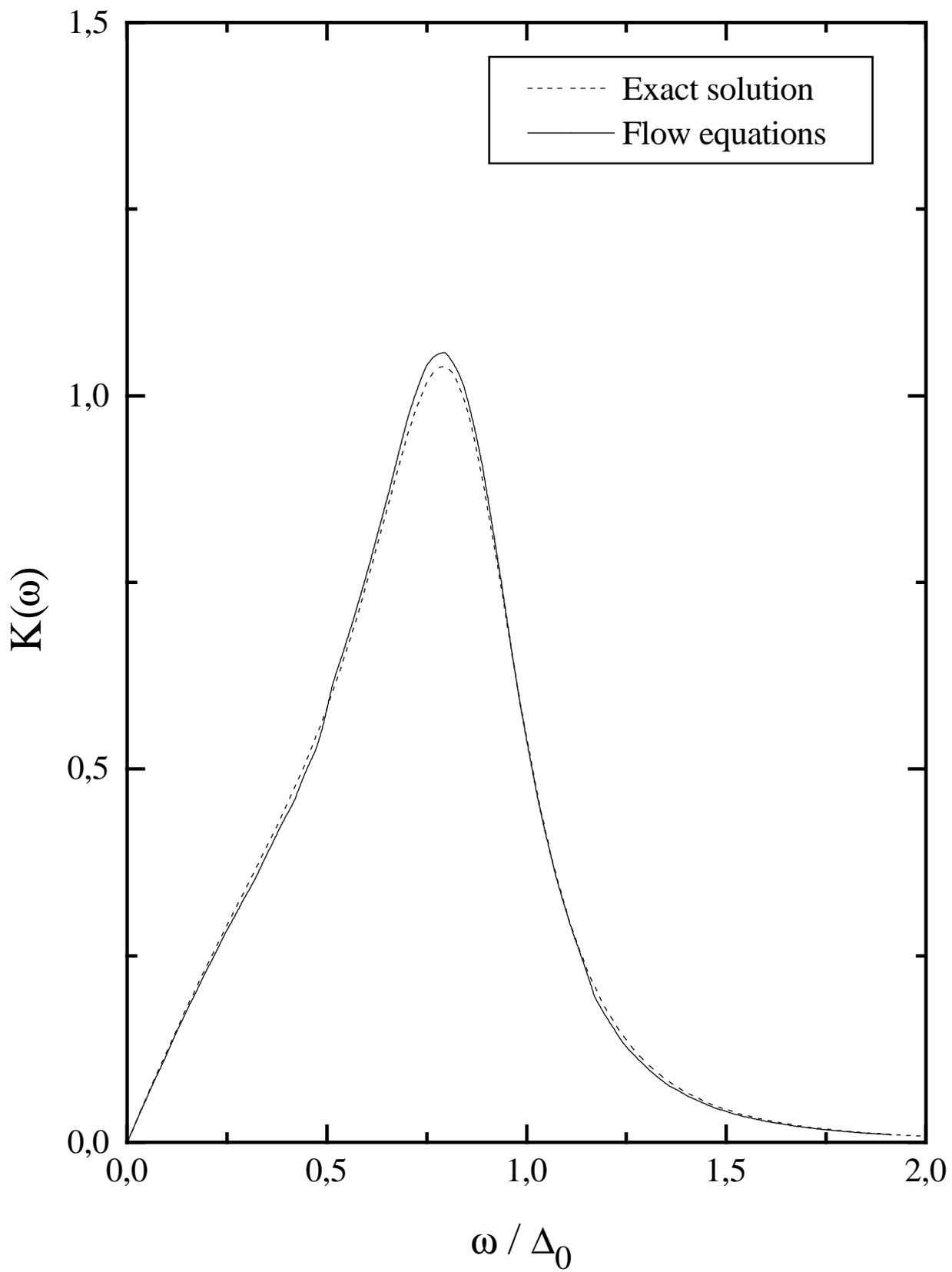

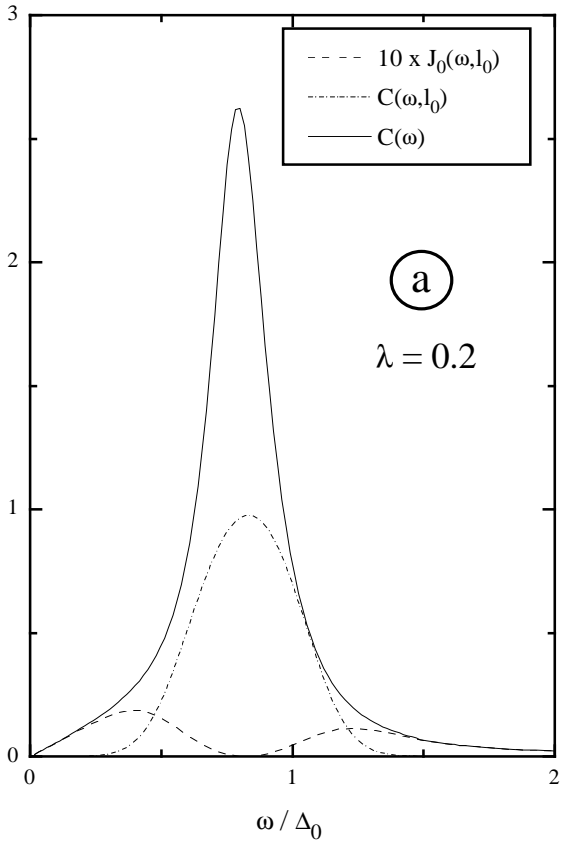
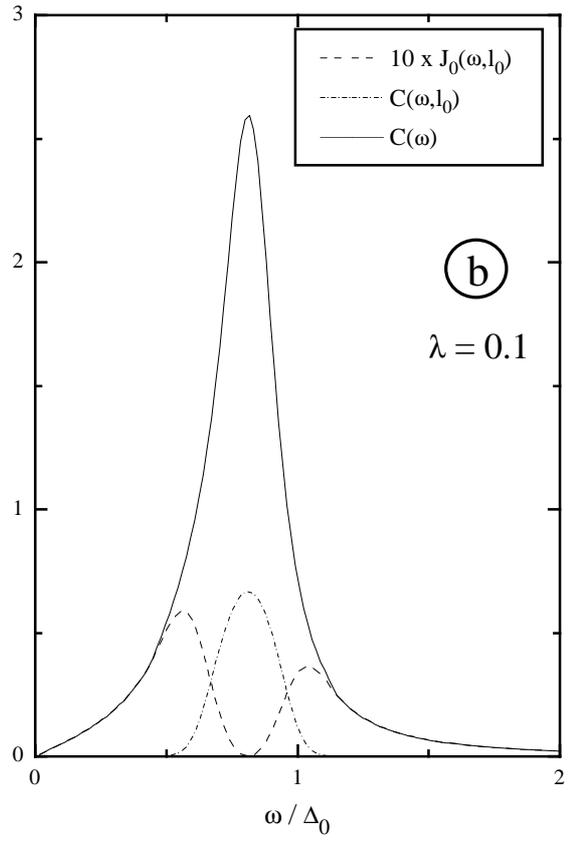
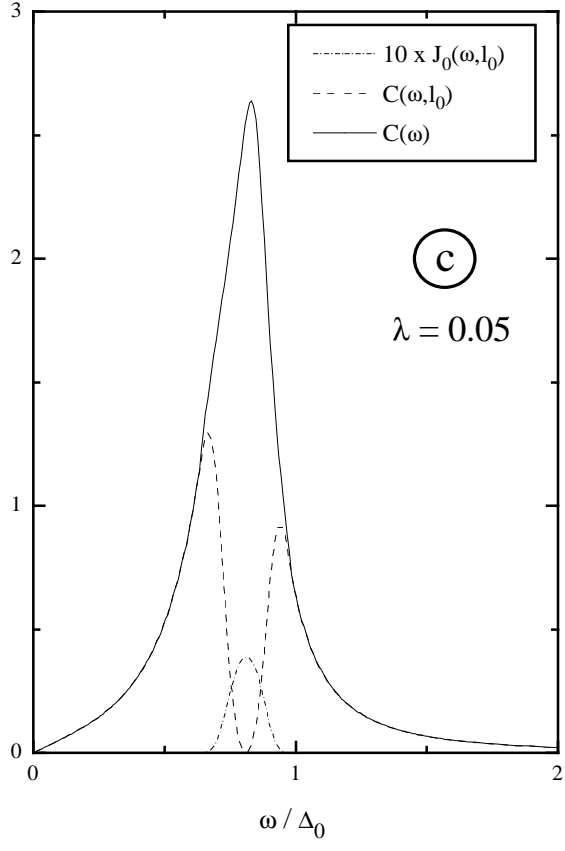
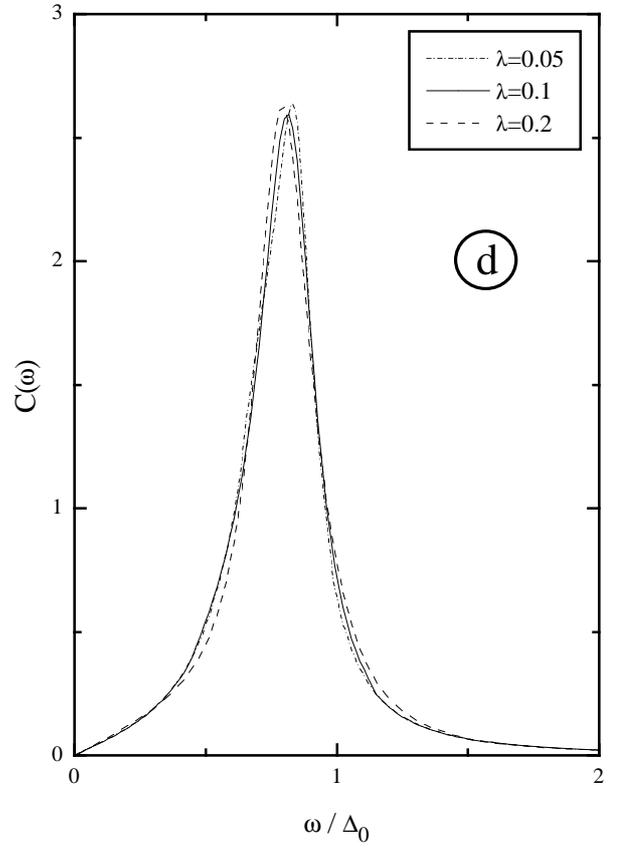

Figure 2

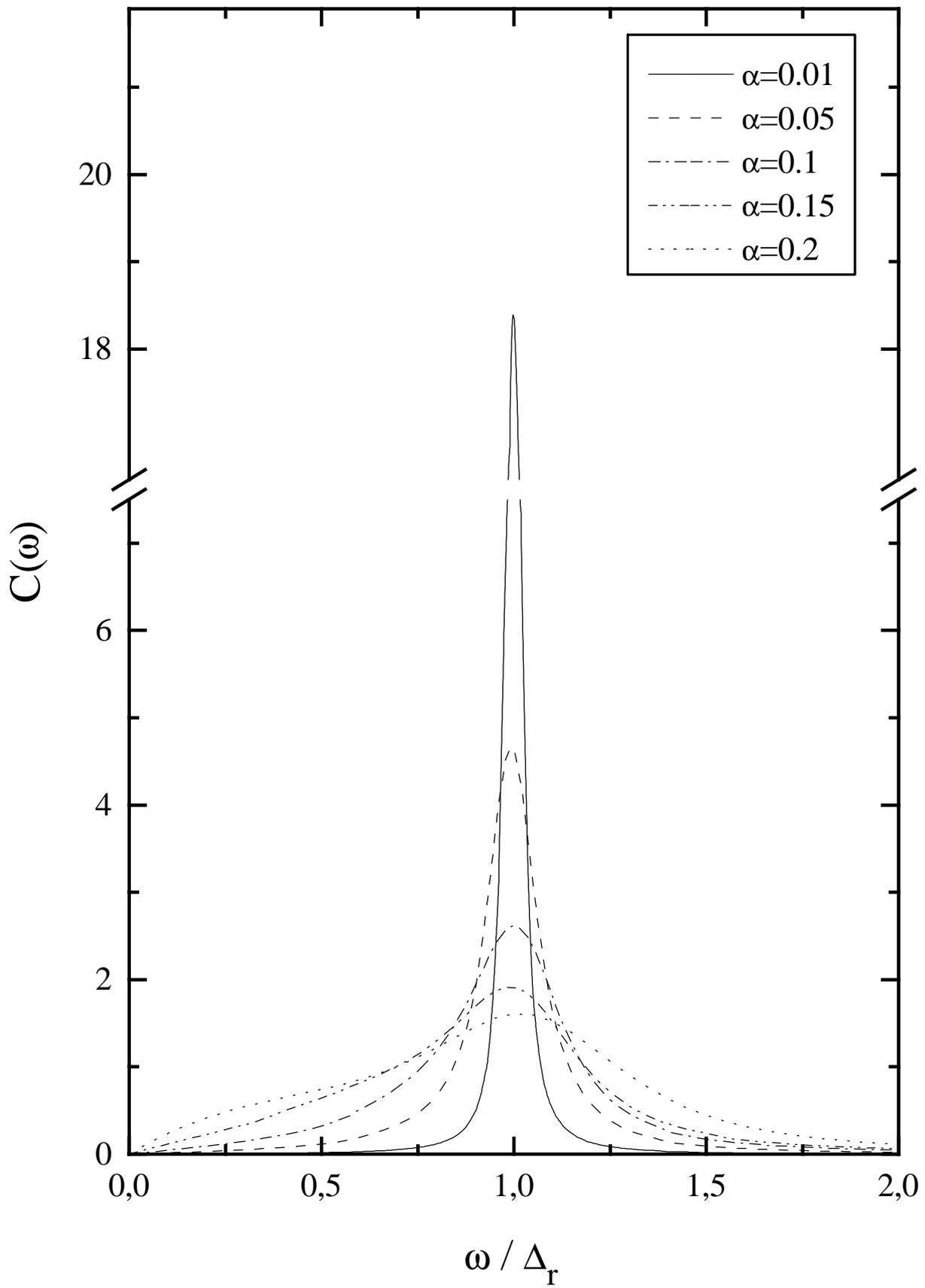

Figure 3

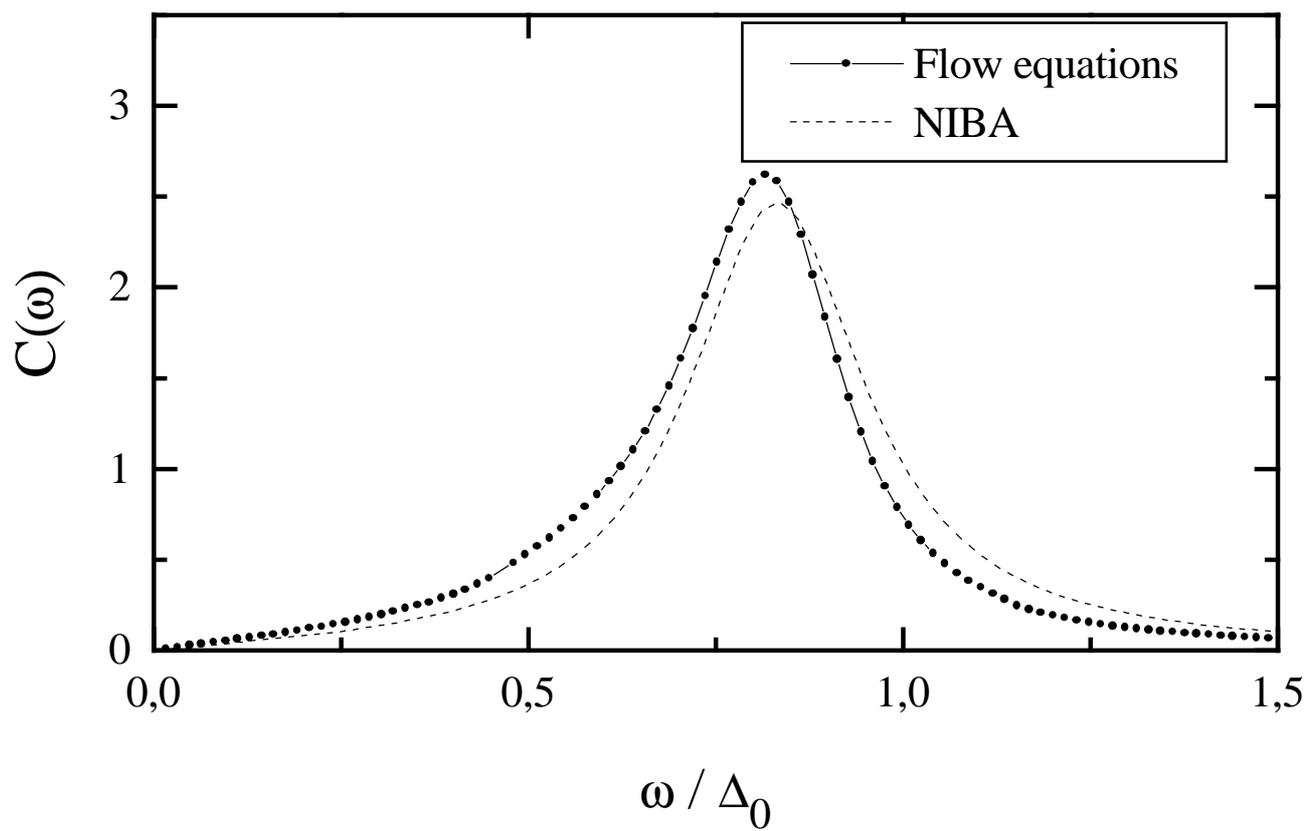

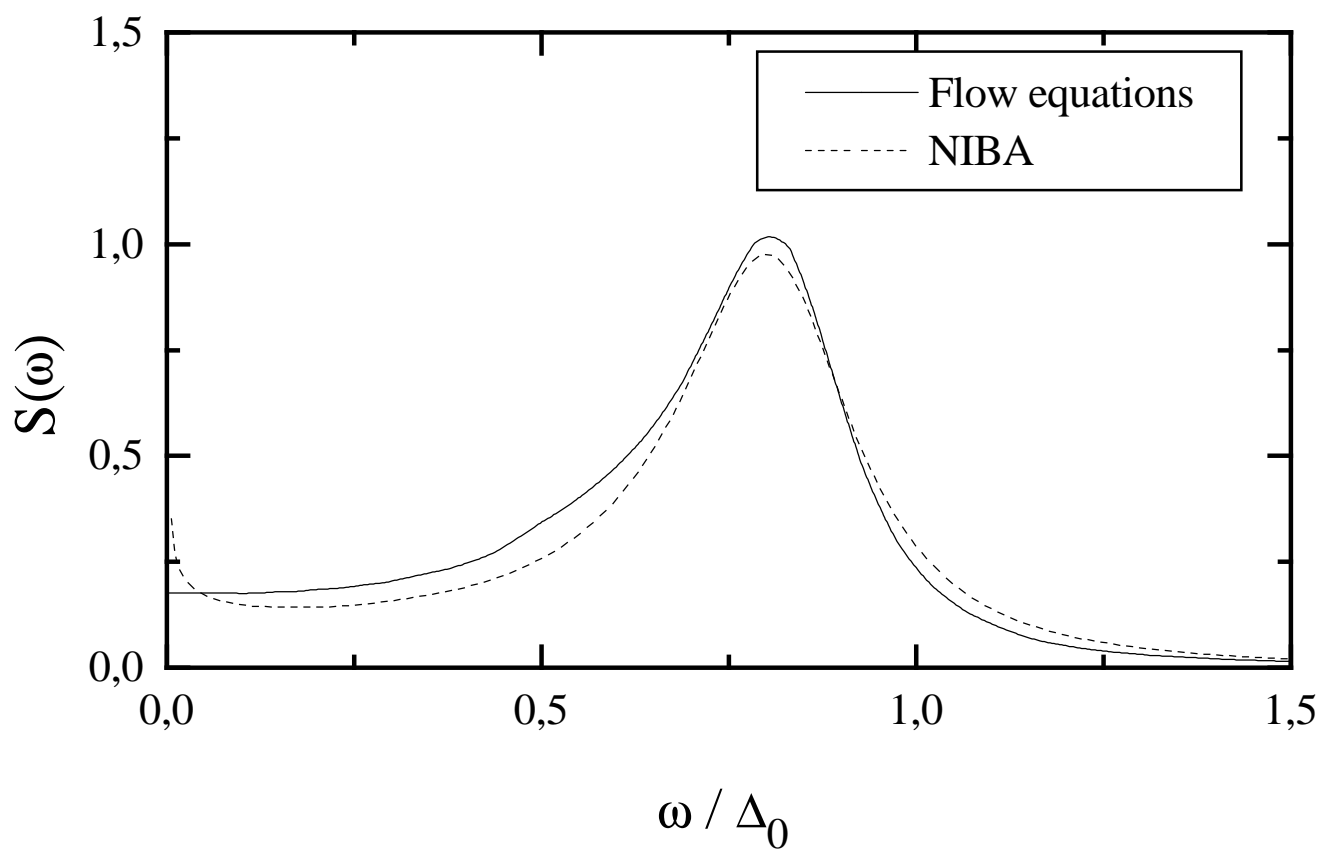

Figure 4

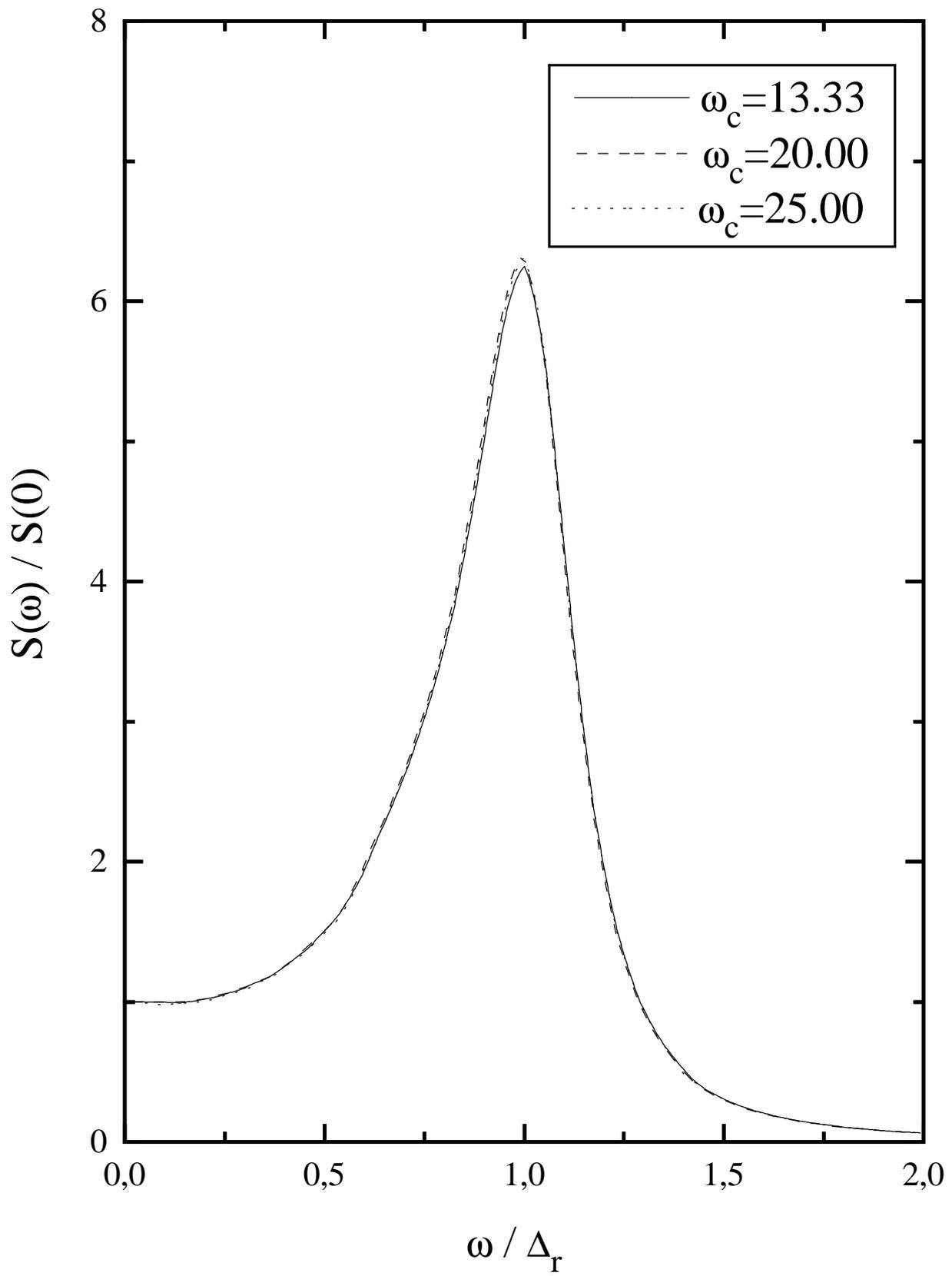

Figure 5

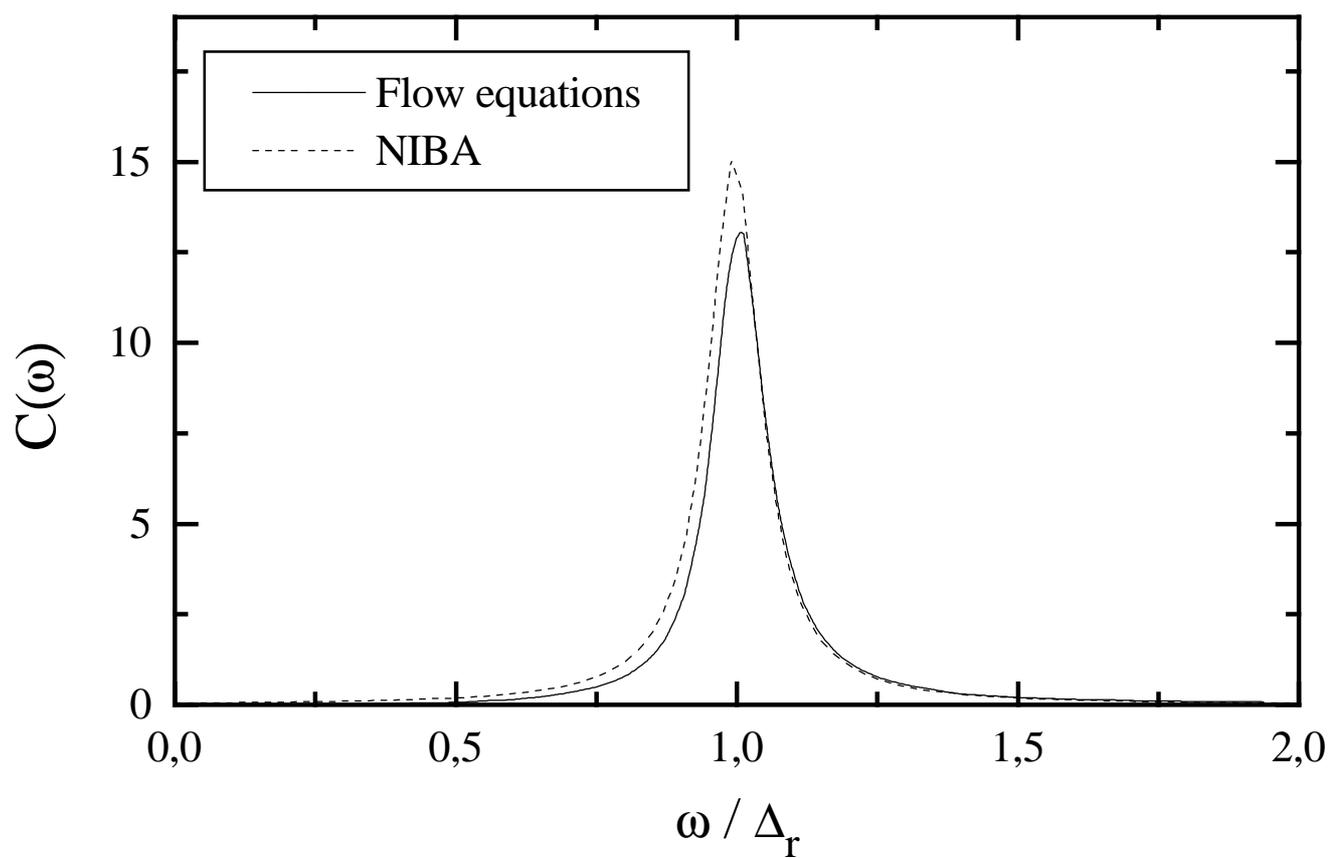
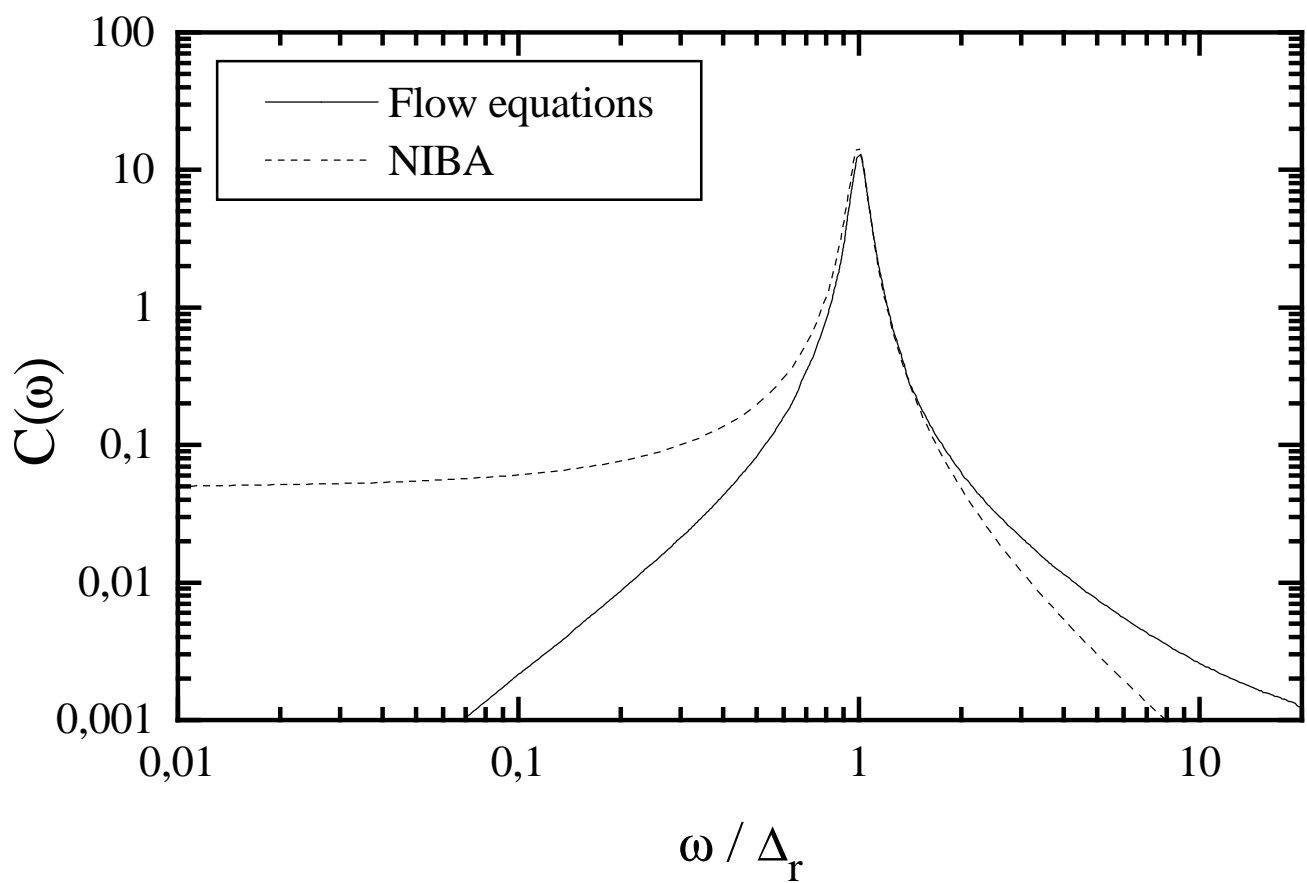

Figure 6

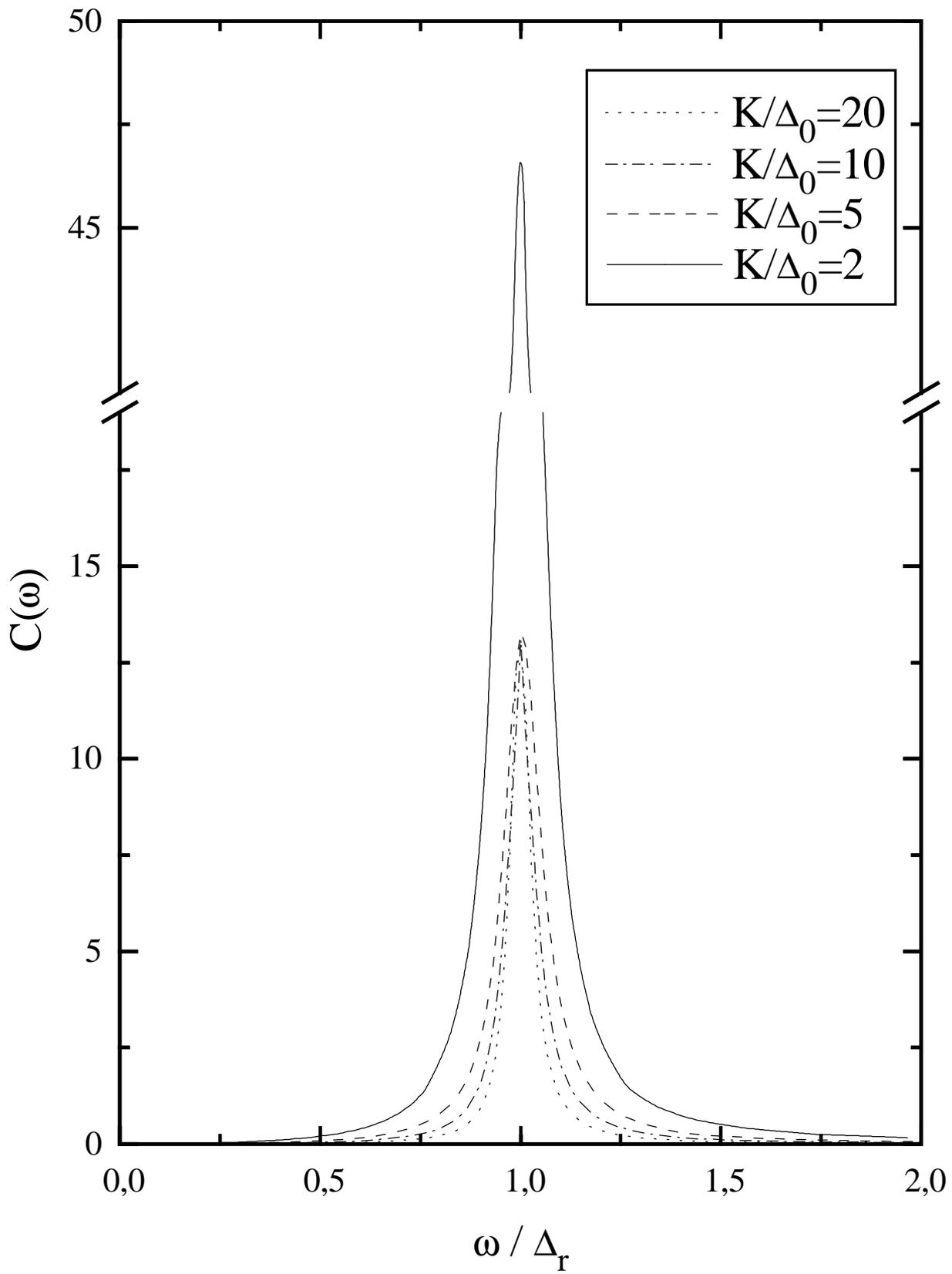

Figure 7